\documentclass[aip,jcp,floats,amsmath,amssymb,reprint,floatfix]{revtex4-1}
\usepackage{graphicx}
\usepackage{dcolumn}
\usepackage{bm}

\usepackage{hyperref}
\usepackage{microtype}

\usepackage{algpseudocode}
\begin{document}


\title{A practical approach to the sensitivity analysis for kinetic Monte Carlo simulation of heterogeneous catalysis}

\author{Max J. Hoffmann}
\affiliation{SUNCAT Center for Interface Science and Catalysis, SLAC National Accelerator Laboratory, Menlo Park, California 94025, United States}
\affiliation{Department of Chemical Engineering, Stanford University, Stanford, California 94305, United States}
\email{maxjh@stanford.edu}

\author{Felix Engelmann}
\affiliation{Chair of Theoretical Chemistry, Technische Universit\"at M\"unchen, Lichtenbergstra{\ss}e 4, 85747 Garching, Germany}
\email{felix.engelmann@eurecom.fr}
\altaffiliation[Now at]{ Department for Communication Systems, EURECOM, 450 Route des Chappes, 06410 Biot, France}

\author{Sebastian Matera}
\affiliation{Institute for Mathematics, Freie Universit\"at Berlin, Arnimallee 6, 14195 Berlin, Germany}
\email{matera@math.fu-berlin.de}

\begin{abstract}
Lattice kinetic Monte Carlo simulations have become a vital tool for predictive quality atomistic understanding of complex surface chemical reaction kinetics over a wide range of reaction conditions. In order to expand their practical value in terms of giving guidelines for atomic level design of catalytic systems, it is very desirable to readily evaluate a sensitivity analysis for a given model. The result of such a sensitivity analysis quantitatively expresses the dependency of the turnover frequency, being the main output variable, on the rate constants entering the model. In the past the application of sensitivity analysis, such as Degree of Rate Control, has been hampered by its exuberant computational effort required to accurately sample numerical derivatives of a property that is obtained from a stochastic simulation method. In this study we present an efficient and robust three stage approach that is capable of reliably evaluating the sensitivity measures for stiff microkinetic models as we demonstrate using CO oxidation on Ru${\rm{O}_{2}}$(110) as a prototypical reaction.
In a first step, we utilize the Fisher Information Matrix for filtering out elementary processes which only yield negligible sensitivity. Then we employ an estimator based on linear response theory for calculating the sensitivity measure for non-critical conditions which covers the majority of cases. Finally we adopt a method for sampling coupled finite differences for evaluating the sensitivity measure of lattice based models. This allows efficient evaluation even in critical regions near a second order phase transition that are hitherto difficult to control. The combined approach leads to significant computational savings over straightforward numerical derivatives and should aid in accelerating the nano scale design of heterogeneous catalysts.
\end{abstract}


\maketitle

\section{Introduction}
The last few years have seen tremendous progress in modeling and analyzing heterogeneous catalysis using the first-principles kinetic Monte Carlo approach(1p-kMC)\cite{ISI:000312170100027}. The appealing features of the approach are an elementary reaction mechanism (and corresponding rate constants), which has been  derived from predictive quality electronic structure methods, and a subsequent solution of the resulting master equation by trajectory sampling. The last step thereby only introduces a numerical (tunable) error, in contrast to the prevalent phenomenological kinetics where the employed mean-field approximation might miss qualitative features\cite{temel2007does, matera2011adlayer}. However, kinetic Monte Carlo has some significant drawbacks. Besides the need to perform one reaction event after the other and the concomitant large computational costs, it is a huge effort to obtain parameter sensitivities, which are a measure for the rate-determining steps\cite{campbell1994future,goldbook}. The evaluation is generally costly because sensitivities are not straightforward expectation values such as coverages or rates.

During the last years, a number of approaches for sensitivity  analysis of kMC models have been developed, mostly in the context of biological ``reaction'' networks and the Chemical Master Equation (in this context kMC is often termed stochastic simulation). These can be grouped into two sets of approaches: i) targeting at a reduction of the noise in finite difference approximations\cite{Anderson12p2237,Arampatzis2014,Plyasunov2007724,srivastava2013comparison}, and ii) targeting a direct estimation of sensitivities from the analysis of the simulated trajectories\cite{Cao701341,jcp_1.3690092,jcp1.4905957}. Albeit being improvements to the brute force numerical differentiation, both groups have their limitations. Direct estimation approaches allow obtaining all sensitivities from the same set of trajectories, (for ergodic stationary processes from a single trajectory). On the other hand, the variance of the corresponding estimators increases with increasing time horizon\cite{Anderson12p2237,jcp_1.3690092}. Especially for stiff problems, the later point is a major concern, as it severely affects the estimation of steady state sensitivities, and there are quite some efforts on sampling strategies for stationary sensitivities\cite{jcp_1.3690092,jcp1.4905957}. This problem is
not as severe for the finite difference based approaches. These, however, require additional simulations for every targeted sensitivity and generally a good guess for the difference parameter\cite{McGill20127170}.

For lattice based models as they appear in heterogeneous catalysis, the methodology for sensitivity analysis is not that mature and there is only a limited number of studies addressing these high-dimensional, usually very stiff problems\cite{Drews01112003,Raimondeau03p1174,meskine2009examination,Pantazis13p054115,Arampatzis2014}.

In this study, we device a three-stage strategy	for the estimation of stationary sensitivities, which is suitable for these problems.  We illustrate this strategy on the model for the CO oxidation on RuO$_2$(110) by Reuter\cite{reuter2006first}, which is a popular fruit-fly test problem\cite{herschlag2015,Gelss2016}. We revisit the reaction conditions studied in ref. \cite{meskine2009examination} and can therefore concentrate on discussing the peculiarities of the sensitivity estimation.

The basic idea of the approach is to first try to directly sample the sensitivities from a single trajectory and only to employ the Coupled Finite Differences method\cite{Anderson12p2237} for those sensitivities, for which the direct sampling estimator shows a too high variance. In practice, directly sampling all sensitivities without prior knowledge might soon become unpractical. We therefore first estimate bounds for the sensitivities using the Pathwise Relative Entropy method\cite{Pantazis13p054115} (section \ref{sec:Bounds}). This also allows to extract an expansion parameter, which is needed for our direct sampling approach and represents the system's memory. So, we avoid its estimation in the direct sampling step and can significantly reduce the sampling effort there. Our direct sampling approach is based on a series expansion of integrated linear response functions and presented in section \ref{sec:LinearResponse}. As it turns out, the derived estimator is  very similar to that by Chen and Cao\cite{Cao633827,Cao701341}. However, our estimator is based on a very different reasoning and avoids the use of random time steps. In the end, these two steps allow to significantly reduce the number of required Coupled Finite Differences (CFD) estimates as presented in section \ref{sec:CoupledFiniteDifferences}.

With the addition of the direct sampling, our approach can be viewed as an extension of the approach put forward  by Arampatzis, Katsoulakis, and Pantazis in the context of the chemical master equation\cite{10.1371/journal.pone.0130825}. We find this additional step to be beneficial, as for most of the considered reaction conditions no or just very few CFD estimates remained necessary.

\section{Background}\label{sec:General}
Sensitivity analysis aims at identifying the most important input parameters for a computational model. That is, if we change an input parameter, how much does the model output change. This information can then be used to identify those parameters, which need to be determined more accurately in order to arrive at a more reliable model. Or, we can employ it to find out, which aspects of the underlying physical system could be optimized in order to arrive at a better performance. In first order, such sensitivity information is provided by partial derivatives of the model output with respect to the input parameter. 

In microkinetic modelling of heterogenous catalysis, the central model outputs are the (stationary) reaction rates of one or more target reactions. As the input parameters have an atomistic meaning, sensitivity analysis is a tool to determine which atomistic aspects of the catalyst control the macroscopic reactivity.

In this section, we outline the background for performing sensitivity analysis for 1p-kMC models. We start with the Markov jump process description of kinetics on lattices. We then introduce sensitivity analysis in this context. After introducing the kinetic Monte Carlo methodology as a numerical tool for simulating  Markov jump processes, we detail the our test problem, the CO oxidation on RuO$_2$(110).

\subsection{Master equation}
In chemical kinetics the rare event dynamics, typical for surface catalytic  processes, is exploited by considering a time evolution that is coarse-grained to  the discrete elementary processes of the reaction mechanism. In this description a state $j$ of the system corresponds to a meta-stable state or domain in the microscopic evolution. That is, the time which is spend between two transitions between these domains is large compared to the time spend for one transition. We can then safely assume that in the time between such rare events the system loses all memory of the past by thermal fluctuations and the sequence of states can then be regarded as a Markov jump process. The probabilities $P_i$ of finding the system in state $i$ then obey a Master equation\cite{reuter2012first,jansen2012introduction}
\begin{equation}
\frac{d}{dt} P(t) =  \Gamma P(t)
\end{equation}
where $P(t)=(P_1(t),P_2(t),\ldots)^T$ is the vector of the probabilities. The stochastic generator  $\Gamma$ has the matrix elements
\begin{equation}
\Gamma_{ij}=w_{ij} -\delta_{ij} \sum\limits_l w_{li}
\end{equation}
where $w_{ij}$ is the transition rate for the event $j \rightarrow i$, with $w_{ii}=0$. The (negative) diagonal element $w^{{\rm acc.}}_i=\sum\limits_l w_{li}$  is called the accumulated rate (for state $i$), which is simply the rate for escaping the state $i$. In the following, we consider  processes, which relax to a single stationary distribution $P_{\rm stat.}$. Further, we assume that the transition probability $P_{\rm kMC}(j|i)=w_{ji}/ w^{{\rm acc.}}_i$ defines a discrete time Markov chain, which also relaxes to a single stationary distribution. For these prerequisites, we can proof the convergence of the series expansion employed in sections \ref{sec:Bounds} and \ref{sec:LinearResponse}\cite{suppmaterialAIP}.

In the context of catalysis, the state $i$ can be regarded as an integer vector, which carries the information, which species is adsorbed on which adsorption site at the surface. Due to symmetries, like translational invariance, and the locality of processes, we can group the allowed events $i \rightarrow j$ ($w_{ij}>0$) into different subsets, each assigned to one elementary step (or reaction). This decomposition can be made a partition of the set of all allowed events, i.e. every allowed event is in a single set $\alpha$. For convenience, we impose that all processes $j\rightarrow i$ belonging to the same subset $\alpha$ have the same value for their transition rate. Thus for each reaction $\alpha$, we can define the partial transition matrix $w^\alpha_{ij}$ with the elements
\begin{equation}\label{eq:PartialTransitionRate}
w^\alpha_{ij}=\begin{cases}
              k^\alpha,~ & {\rm if}~  j\rightarrow i \in \alpha,\\
              0,&{\rm otherwise},
             \end{cases}
\end{equation}
where $k^\alpha$ will be called the rate constant (RC) of the reaction $\alpha$ in the following. With $w^\alpha_{ij}$ we can define the (partial) generator $\Gamma^\alpha$ as
\begin{equation}\label{eq:PartialGenerator}
\Gamma^\alpha_{ij}=w^\alpha_{ij} -\delta_{ij} \sum\limits_l w^\alpha_{li}.
\end{equation}
Summing over all (partial) generators, we arrive at the (total) generator $\Gamma$
\begin{equation}\label{eq:SumPartialGenerator}
\Gamma_{ij}=\sum\limits_{ \alpha=1}^{N_{\rm reac.}} \Gamma^\alpha_{ij},
\end{equation}
where $N_{\rm reac.}$ is the total number of reactions we have for our system.
As for the total generator, the absolutes of diagonal elements of the partial generator  are the partial accumulated rate $w^{\alpha,{\rm acc.}}_i=\sum\limits_l w^\alpha_{li}$.

\subsection{Sensitivity analysis}
The central objective of kinetics in heterogeneous catalysis are average, stationary reaction rates $\langle R \rangle$, also called turnover frequencies (TOF), if suitably normalized. In the case, that there are multiple elementary reactions contributing to the targeted overall reaction, $\langle R \rangle$ is the superpositions of the corresponding reaction rates $\langle R^\alpha \rangle$ of the elementary reactions
\begin{equation}\label{eq:ReactionRate}
 \langle R \rangle=\sum\limits_\alpha^{N_{\rm reac.}}T^\alpha \langle R^\alpha \rangle, \text{ with
} \langle R^\alpha \rangle=\sum\limits_{i} w^{\alpha,{\rm
acc.}}_i P_{i,\rm stat.}
\end{equation}
where $P_{i,\rm stat.}$ is the stationary probability distribution and $T^\alpha$ is a constant which depends only on the stoichiometry of the reaction network. For the below discussed CO oxidation on RuO$_2$(110), there are four reactions which yield one CO$_2$ molecule per event. In this case, 
$T^\alpha$ for the CO oxidation rate is one for each of these four and zero for all other reactions.

We require a unique stationary probability distribution $P_{\rm stat.}$ and the reaction rate $\langle R \rangle$ is therefore a function of the rate constants. As the RCs measure how fast the elementary reactions proceed, finding out how $\langle R \rangle$ reacts on little changes in the RCs, can be used to identify the most important steps, i.e. those steps which need to be accelerated or slowed down to achieve a higher reaction rate. In chemical kinetics, a useful measure for such sensitivity analysis is\cite{meskine2009examination}
\begin{equation}\label{eq:DRC}
 X^{\alpha}:= \frac{k^\alpha}{{ \langle R \rangle}} \left(\frac{\partial \langle R
\rangle (\{k^\beta\})}{\partial k^\alpha} \right)_{k^{\phi\neq \alpha}}.
\end{equation}
i.e. the relative change of $\langle R \rangle$ over the relative change of the corresponding RC, while all other RCs are kept constant. We will term this sensitivity index Degree of Rate Sensitivity (DRS). 

The DRS can be regarded as generalized reaction order, which becomes clear for the case that all DRS do not change for a range of values for the RCs.
In this range, the reaction rate obeys the relation
\begin{equation}\label{eq:GeneralizedReactionOrder}
 \langle R \rangle= r_0 \prod\limits_{\alpha}^{N_{\rm reac.}} (k^\alpha)^{X^\alpha}
\end{equation}
where $r_0$ is independent of the RCs. For elementary steps involving gas phase species $\rm A$ like adsorption or Eley-Rideal reactions, we expect rate expressions of the form
$k^\alpha\propto p_{\rm A}$ with the partial pressure $p_{\rm A}$. We then arrive at a power law kinetics
\begin{equation}\label{eq:ReactionOrder}
 \langle R \rangle=\tilde{r}_0  \prod\limits_{\rm A} (p_{\rm A})^{\nu_{\rm A}},\quad \text{with}~ \nu_{\rm A}=\sum\limits_{\alpha \in R_{\rm A}} X^\alpha
\end{equation}
where $R_{\rm A}$ is the set of elementary steps involving the gas phase species $\rm A$ and $\tilde{r}_0$ is independent of the partial pressures. Thus the DRS connect 
macroscopic reaction orders with microscopic elementary steps and rate constants. Usually, we employ the rate expressions $k^\alpha=f^\alpha(T,\{p_{\rm A}\}) \exp (-\Delta E^\alpha/k_{\text{B}}T)$, where $k_{\text{B}}$ is the Boltzmann constant, $T$ is the temperature and $\Delta E^\alpha$ is the activation barrier for the reaction. Using sensitivity analysis, we can find a clear connection between the elementary barriers and the macroscopic apparent activation barrier $E_{\rm app.}$ by\cite{meskine2009examination}
\begin{equation}\label{eq:ApparentActivation}
 E_{\rm app.}:= - \left(\frac{\partial \langle R \rangle (T,\{p_{\rm A} \})}{\partial~ (k_BT)^{-1}}\right)_{p_{\rm A}} \approx \sum\limits_{\alpha}^{N_{\rm reac.}} X^\alpha \Delta E^\alpha.
\end{equation}
For brevity, we have provided only the dominant contribution and neglected the (typically weak) temperature dependence of the pre-exponential factor.

To simplify the later discussion, we decompose the  DRS
\begin{equation}
\frac{k_i}{\langle R \rangle}\left(\frac{\partial \langle R \rangle (\{k^\beta\})}{\partial k^\alpha} \right)_{k^{\beta\neq \alpha}}= X^{\alpha}_0 + X^{\alpha}_1
\end{equation}
using the product rule. Here  $X^{\alpha}_0=T^\alpha \langle R^\alpha\rangle/\langle R \rangle$, which is the relative contribution to the total rate of the reaction $\alpha$. As this is  a stationary expectation, it is straightforward to obtain from sampling. Therefore we  will concentrate on the second term
\begin{equation}
 X^{\alpha}_1=\frac{k_i}{\langle R \rangle}\sum\limits_i R_i \frac{\partial
P_{i,{\rm stat.}}}{\partial k^\alpha}
\end{equation}
in the further discussion. Intuitively, $X^{\alpha}_1$ can be regarded as the average effect on the creation and annihilation of states $i$, which allow for those reactions contributing
to the overall reactivity.
As $X^{\alpha}_1$ originates from the $k^\alpha$-dependence of the probability distribution, this can not be expressed as a stationary expectation, except for the case when the stationary distribution is known as a function of the RCs.  For obtaining $X^{\alpha}_1$, we can omit any explicit dependence of $R$ on variations of the RCs.

The DRS is strongly related to the Campbell's Degree of Rate Control (DRC)\cite{campbell1994future}
\begin{equation}\label{eq:DRCCamp}
 X^{\xi}_{\rm DRC}=\frac{k^\xi_+}{{ \langle R \rangle}} \left(\frac{\partial \langle R
\rangle (\{k_+^{\phi}\}, \{K^{\phi}\})}{\partial k^\xi_+} \right)_{k_+^{\phi\neq \xi}, K^\phi}
\end{equation}
where $\xi$ denotes a pair of forward and backward reaction and $k^\xi_+$ and $K^\xi$ are the corresponding RCs of the forward reaction and the equilibrium constant, respectively. 
The only difference between DRC and DRS is what is kept constant during differentiation, all other RCs (DRS) or all other forward RCs and the equilibrium constants (DRC). In a previous publication we therefore termed both Degree of Rate Control\cite{meskine2009examination}. To clarify the presentation, we have introduced a new name for the less common sensitivity measure \ref{eq:DRC}.
Applying the chain rule, we arrive at a simple relation between DRS $X^{\alpha}$ and DRC $ X^{\xi}_{\rm DRC}$\cite{meskine2009examination}
\begin{equation}\label{eq:WhyWeShouldAlwaysCalculateOurDRCs}
 X^{\xi}_{\rm DRC}=X^{\xi_+} + X^{\xi_-} 
\end{equation}
where $\xi_+$ identifies the forward reaction and $\xi_-$ the backward reaction. The DRC $ X^{\xi}_{\rm DRC}$ is therefore  easily obtained by adding the DRSs for the forward and the reverse reaction. 

By construction, the DRC $ X^{\xi}_{\rm Camp.}$ is zero for reactions which are in equilibrium. Identifying rate-determining steps by large $| X^{\xi}_{\rm DRC}|$ agrees then with chemical intuition, which would not assign the property 'rate-determining' to steps which are in equilibrium. This thermodynamic consistency is absent in the definition \ref{eq:DRC} for the $X^{\alpha}$. On the other hand, the DRSs $X^{\alpha}$ provide the sensitivity to all input parameters and thereby also allow to judge on the impact of equilibrated reactions. Further, there exists no equivalent to  Eqn. \ref{eq:GeneralizedReactionOrder} to \ref{eq:ApparentActivation} for the DRC. For a more detailed discussion of the microscopic meaning and the relation between them we refer the reader to reference\cite{meskine2009examination}.

Which of both measures is easier to interpret or provides a deeper insight, is primarily a matter of taste. Most information can probably be extracted using both and, by Equation \ref{eq:WhyWeShouldAlwaysCalculateOurDRCs}, there is no limiting factor preventing us from doing so.
We will, however, concentrate on the DRS for most of the manuscript for the mere reason of shorter equations and a more comprehensive and compact presentation of the theoretical background.
We come back to the DRC when we discuss our final results in terms surface coverage and barriers.

\subsection{Kinetic Monte Carlo}
In most cases, the Master equation is too high-dimensional to be solved directly. Instead, one will simulate  the underlying Markov process and obtain estimates of the targeted expectations by averaging over trajectories. For this and all our implementations we use the  {\sc kmos} code package\cite{Hoffmann20142138}, which we have developed during the last years for lattice based kinetic Monte Carlo.

We employ the so-called Variable-Step Size Method\cite{jansen2012introduction}, which can be described as follows
\begin{enumerate}
 \item Initialize the state $i$ and the time $t$\label{kMC:step1}
 \item Increase the time $t \rightarrow t + \delta t$ with $\delta t = (w^{\rm acc.}_i)^{-1} \log \xi$, where $\xi$ is a uniformly distributed random variable on $(0,1]$.
\item Choose the event $i \rightarrow j $ with the probability $P_{\rm kMC}(j|i)= w_{ji}/w^{\rm acc.}_i$ \item Set $i=j$ and go to \ref{kMC:step1} or stop if the termination criterion is reached.
\end{enumerate}
The method essentially simulates a classical discrete time  Markov chain with the transition probability matrix $ P_{\rm kMC}$ and simply adjusts the time steps between the jumps to arrive at a statistically proper continuous time Markov jumps process.

Focusing on stationary reaction rates, we can obtain estimates most conveniently by time averaging. That is, we will approximate $\langle R^\alpha \rangle$, after a sufficiently long relaxation, with
\begin{equation}\label{eq:StatkMCAv}
 \langle R^\alpha \rangle \approx \frac{1}{\tau_N}\sum\limits_{n=0}^N w^{\alpha,{\rm
acc.}}_{i_n} \delta t_n \approx \frac{1}{t_N}\sum\limits_{n=0}^N w^{\alpha,{\rm
acc.}}_{i_n} \Delta t_n
\end{equation}
where $( i_1,\ldots ,i_N)$ is the sequence of states generated by the kMC simulation and $(\delta t_1, \ldots ,t_N)$ are the corresponding time steps with $\tau_N=\sum\limits_{n=0}^N\delta t_n$. For the second approximation we have pre-averaged the time steps, i.e. $\Delta t_n= (w^{\rm acc.}_{i_n})^{-1}$ and $t_N=\sum\limits_{n=0}^N\Delta t_n$. This is a common approximation\cite{jcp_1.3690092,jcp1.4905957} to reduce the noise caused by the random time steps and the introduced bias vanishes for large enough $N$. 

In practice, we will employ a number independent sample trajectories to estimate an expectation $\langle A \rangle\approx \bar{A} = K^{-1} \sum\limits_k A_k$ and the standard deviation of our estimate $\sigma(\bar{A})^2\approx (K^2-K)^{-1} \sum\limits_k (A_k -  \bar{A})^2$. Here $K$ is number of samples and ${A}_k$ are time averages. We generate these samples by first simulating the one trajectory. We then perform a number of kMC steps for decorrelation and then use the last configuration as initial condition for the next trajectory. We repeat this until we have reached the total number $K$ of samples.

\subsection{CO oxidation on RuO$_2$(110)}\label{sub:COoxidation}
Throughout this article, we consider the realistic first-principles kinetic Monte Carlo (1p-kMC) model for the CO oxidation on RuO$_2$(110) by Reuter\cite{reuter2006first}. It is one of the most thoroughly tested and understood 1p-kMC models (e.g. the refs. \cite{matera2011adlayer,herschlag2015,liu2015transitions}) for which we have already performed a sensitivity analysis\cite{meskine2009examination}. This makes it well-suited for testing our scheme and we can concentrate on discussing the peculiarities of the employed sampling approaches.

In detail, the model considers two kinds of adsorption sites, the so-called bridge (br) and coordinatively un-saturated (cus) sites. These are arranged on a square lattice with alternating rows. Oxygen and CO can adsorb on both kind of sites, where oxygen adsorbs dissociatively while CO stays intact after adsorption. CO$_2$ does not bind to the surface and the sites can be in three different states: i) empty, ii) CO occupied and ii) atomic oxygen occupied. Altogether, the model comprises 22 elementary  reaction steps, which are single site CO ad/desorptions, nearest-neighbors O$_2$  ad/desorptions, adsorbed CO/O nearest-neighbor diffusion, and formation of gaseous CO$_2$ by Langmuir-Hinshelwood reactions of co-adsorbed CO and O. The corresponding RCs have been obtained from Density Functional Theory using expressions based on harmonic Transition State Theory (for details see  refs. \cite{reuter2006first,reuter2012first}). By these expressions, the rate constants depend on the temperature $T$ as well as on the CO and O$_2$ partial pressures $p_{\text{CO}}$ and $p_{\rm O_2}$, respectively. We will consider the reactions conditions  of $T=600$\,K, $p_{\text{O}_2}=1$\,bar, and varying $p_{\text{CO}} \in [0.05,50]$\,bar, which result in the rate constants given in table \ref{table1}. These are the reactions conditions, which we have employed in our previous studies\cite{meskine2009examination,matera2011adlayer} and which therefore allow for direct comparison.

\begin{table}
\caption{\label{table1}
Elementary reaction steps and corresponding rate constants at the considered reaction conditions of $T=600$\,K, $p_{\text{O}_2}=1$\,bar, and varying $p_{\text{CO}} \in [0.05,50]$\,bar. An index $_{cus}$ or $_{br}$ indicates that the molecule is adsorbed at a cus- or br-site, respectively.}
\begin{tabular}{l|c|c|}
Process							& $\Delta E^\alpha$\,(eV)	&   rate constant (s$^{-1}$) 			\\[0.1cm] \hline
Adsorption						& 	&    					\\[0.1cm]
$ {\rm CO } \rightarrow {\rm CO_{cus}}$			& 0.0	&   $2\times10^8\times p_{\text{CO}} $(bar) 			\\
$ {\rm  CO  } \rightarrow  {\rm CO_{br}}$		& 0.0	&  $2\times10^8\times p_{\text{CO}} $(bar) 			\\
$ {\rm  O_2  } \rightarrow  {\rm O_{cus} + O_{cus}}$	& 0.0	&  $9.7 \times 10^7$			\\
$ {\rm  O_2  } \rightarrow  {\rm O_{br} + O_{br}}$	& 0.0	&   $9.7 \times 10^7$			\\
$ {\rm  O_2  } \rightarrow  {\rm O_{br} + O_{cus}}$	& 0.0	&  $9.7 \times 10^7$			\\[0.1cm]
Desorption						&	&    					\\[0.1cm]
$ {\rm CO_{cus}}  \rightarrow {\rm CO}$			& 1.3	&	$9.2 \times 10^{6}$		\\
$ {\rm  CO_{br}}   \rightarrow  {\rm CO}$		& 1.6	&	$2.8 \times 10^{4}$		\\
${\rm O_{cus} + O_{cus}} \rightarrow  {\rm  O_2  }$ 	& 2.0	&	$ 2.8\times 10^1$		\\
$  {\rm O_{br} + O_{br}} \rightarrow {\rm  O_2  }$	& 4.6	&	$ 4.1\times 10^{-21}$		\\
$  {\rm O_{br} + O_{cus}} \rightarrow {\rm  O_2  }$	& 3.3	&	$ 3.4\times 10^{-10}$		\\[0.1cm]
Diffusion						&	&        				\\[0.1cm]
${\rm CO_{cus} }  \rightarrow  {\rm CO_{cus}}$		& 1.7	&	$6.6	\times 10^{-2}$ 	\\
${\rm CO_{br}  } \rightarrow  {\rm CO_{br}} $		& 0.6	&	$1.1\times 10^{8}$		\\
${\rm CO_{cus} }  \rightarrow  {\rm CO_{br}}$		& 1.3	&	$1.5\times 10^{2}$		\\
${\rm CO_{br} }  \rightarrow  {\rm CO_{cus}}$		& 1.6	&	0.5				\\
${\rm O_{cus} } \rightarrow  {\rm O_{cus}} $		& 1.6	&	0.5				\\
${\rm O_{br}  } \rightarrow  {\rm O_{br}} $		& 0.7	&	$1.6\times 10^{7}$		\\
${\rm O_{cus}  } \rightarrow  {\rm O_{br}} $		& 1.0	&	$4.9\times 10^{4}$		\\
${\rm O_{br}  } \rightarrow  {\rm O_{cus}} $		& 2.3	&	$6.0\times 10^{-7}$		\\[0.1cm]
CO$_2$ formation					&	&       				\\[0.1cm]
$ {\rm CO_{cus} + O_{cus} } \rightarrow  {\rm CO_2}   $	& 0.9	&	$1.7 \times 10^5$		\\
$ {\rm CO_{br}  + O_{br} } \rightarrow  {\rm CO_2}  $	& 1.5	&	1.6				\\
$ {\rm CO_{cus}  + O_{br} } \rightarrow  {\rm CO_2}  $	& 1.2	&	$5.2 \times 10^2$		\\
$  {\rm O_{cus} + CO_{br} } \rightarrow  {\rm CO_2} $	& 0.8	&	$1.2 \times 10^6$		\\ [0.1cm]\hline
\end{tabular}
\end{table}

\begin{figure}
\includegraphics[width=\linewidth]{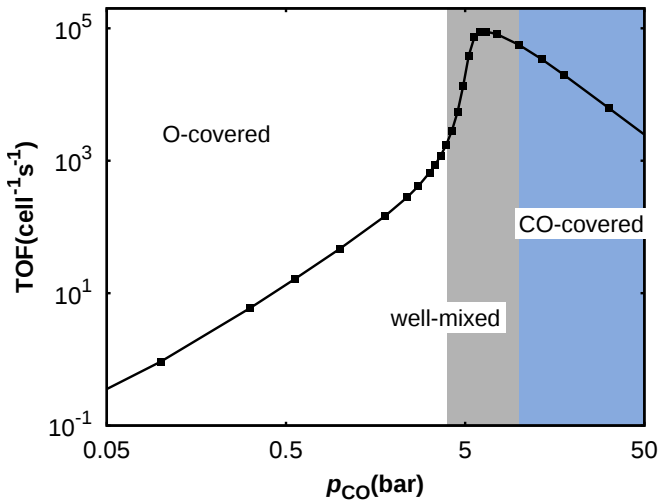}
\caption{Turnover frequency for the CO oxidation on RuO$_2$(110) for  $T$=600\,K, $p_{\text{O}_2}=1$\,bar and varying $p_{\text{CO}}$. At low $p_{\text{CO}}$ the surface is oxygen covered and the TOF is low (white background). With increasing partial pressure $p_{\text{CO}}$ the CO coverage increases as well. The TOF increases having its maximum, when both CO and oxygen are present in appreciable amounts (gray background). At high $p_{\text{CO}}$ the surface becomes CO-covered and the TOF decreases again (blue background).
\label{TOFfig}}
\end{figure}
The TOF for CO oxidation in considered range of reaction conditions is shown in Fig. \ref{TOFfig}, i.e. the sum of the rates for the four CO$_2$ formation reactions normalized to the size of a surface unit cell. The data was obtained employing a lattice with $20 \times 20$ surface unit cells and by averaging over $10^9$ kMC steps, after an initial relaxation of $10^8$ steps. These are save settings, as we observe that stationary operation is usually reached after $10^6$ steps. We find three regimes, which are characterized by their dominant site occupations: i) at low $p_{\text{CO}}$ the surface is almost fully oxygen covered and the reactivity is very low (white background), ii) the TOF is the highest for medium  $p_{\text{CO}}$, where both CO and O occupy the surface sites in roughly the same amounts (grey background), and iii) the TOF decreases for higher $p_{\text{CO}}$ as now the surface becomes CO poisoned (blue background). The reaction rates for all elementary steps can be found in the supplementary material\cite{suppmaterialAIP}.

\section{Methods}
We now turn to the problem of estimating the DRS from kinetic Monte Carlo simulations. As written above, we propose a three-step strategy for that purpose. 

In the first step, we estimate bounds for the DRS using the Relative Entropy Method as presented in section \ref{sec:Bounds}. The ingredients to these bounds are the pathwise Fisher Information Metric and the time-integrated auto-correlation function (TAC) of the reaction rate. Especially, the later is not trivial to estimate and, in section \ref{sub:estimatorTAC},  we present a sampling strategy based on a series expansion of a generalized inverse.

In the second step, we try to directly sample all DRS from a single trajectory. For this, we employ our Linear Response Theory based estimator, which we present in section \ref{sec:LinearResponse}. The bounds from the first step help here to reduce sampling overhead, because we can skip those DRS which bounds are close to zero. We further utilize a numerical parameter, which we need to adjust during for the determination of the TAC. 

The direct estimator might suffer from a too high variance for some sensitivities and we would need an unrealistic increase of the sampling effort to achieve an satisfactorily accuracy. In the third step, we therefore determine these sensitivities using Coupled Finite Differences (CFD) \cite{Anderson12p2237}, which might provide the desired variance with less sampling effort. However, we will need an extra kMC simulation for each DRS and if we can determine the at least some DRS sufficiently from direct sampling, we save computational costs.  We outline how to implement CFD for lattice problems in section \ref{sec:CoupledFiniteDifferences}.

\subsection{Bounds for the Degree of Rate Sensitivity from the Relative Entropy Method}\label{sec:Bounds}
As written above, this step's purpose is to reduce the computational costs at the later stages of the scheme. For this, we estimate upper bounds for all $X^{\alpha}_1$ and we can save CPU time by discarding those $X^{\alpha}_1$ which are close to zero. From this estimation process, we also obtain reasonable truncation limits for next steps in section \ref{sec:LinearResponse}, the direct sampling of $X^{\alpha}_1$ based on time-integrated linear response function.

For the bound, we will follow an approach put forward by Arampatzis, Katsoulakis, and Pantazis\cite{Pantazis13p054115,10.1371/journal.pone.0130825}. For the DRS, their bound for a general parametric dependence specializes  to
\begin{equation}\label{eq:bounds}
 |X^{\alpha}_1| \leq \frac{1}{\langle R \rangle}\sqrt{~ c_R I^{\alpha \alpha}}=:B^\alpha
\end{equation}
where $I^{\alpha \alpha}$ are the diagonal elements of the pathwise Fisher Information Metric (FIM), and $c_R$ is the time-integrated auto-correlation of the reaction rate. When the transition matrices $w^\beta_{ij }(\theta)$ depend arbitrarily on a parameter vector $\theta$, the Fisher Information Metric is given by
\begin{equation}
I=  \sum\limits_{j \rightarrow i} P_{j,{\rm stat.}}w_{ij} \nabla_\theta \log w_{ij} \nabla_\theta \log w_{ij}.
\end{equation}
where the sum runs over all allowed events $j \rightarrow i$ ($w_{ij}\neq 0$). Using Eq. \ref{eq:PartialTransitionRate} and the logarithms of the rate constants as parameters, the above equation reduces to the reaction rates
\begin{equation}\label{eq:FIMandRates}
 I^{\alpha \beta}= \delta_{\alpha\beta} \langle R^\alpha \rangle.
\end{equation}
So, using bound \ref{eq:bounds} will simply sort out those reactions which do not happen frequently enough. In the stationary case, the time-integrated auto-correlation (TAC)  is
\begin{equation}
\begin{split}
 c_R &= 2 \int\limits_0^\infty \langle \delta R(t) \delta R \rangle dt\\
 &= \sum\limits_{ij} 2
\int\limits_0^\infty  \delta R_i(e^{\Gamma t})_{ij}  \delta R_j  P_{j,{\rm stat.}} dt
 \end{split}
\end{equation}
where $\delta R_i= R_i - \langle R \rangle$.

The bound \ref{eq:bounds} can be regarded as a conservative sensitivity measures, i.e. using only the bounds we will definitely not miss an important elementary step. Even only considering the Fisher Information Matrix (FIM) allows for a sensitivity analysis\cite{Pantazis2013}, albeit with a different objective. The FIM measures the impact of a small change in the parameters on the whole stochastic process, i.e. the leading order of the relative entropy between the original and the perturbed process. This allows for a global picture, but if one is interested in  a particular average, the FIM based findings might be misleading. For example, consider the case with two types of sites $a$ and $b$, which are completely decoupled. Further, we want to assume that the reaction takes only place on sites of type $a$, but the processes on site $b$ are much faster. By Eq. \ref{eq:FIMandRates}, the processes on $b$-sites will have the largest FIM $ I^{\alpha \alpha}$ and therefore the largest $B^\alpha$. However, from the construction of our example, we know that these processes can have no impact on the average reaction rate.

\subsubsection{An estimator for the time-integrated auto-correlation based on generalized inverses}\label{sub:estimatorTAC}
With the definitions $v_i= \delta R_i$ and $u_i=\delta R_i  P_{i,{\rm stat.}}$, the TAC can be written as
\begin{equation}\label{eq:TACPseudoInverse}
\begin{split}
 c_R&=2(v,\int\limits_0^\infty e^{\Gamma t} dt  u)=-2(v, \Gamma^\# u)\\
 &=2(v, D \sum\limits_{l=0}^\infty(1 - \Gamma D)^l u)
\end{split}
\end{equation}
where $(\cdot,\cdot)$ is the usual standard scalar product and $\Gamma^\#$ is a generalized inverse. The first line results from $u$ and $v$ being perpendicular to the left ($g_0=(1,1,\cdots ,1)^T$) respectively the right eigenvector ($g^0_i= P_{i,{\rm stat.}}$) to the eigenvalue zero. For the second line, we have employed the Neumann series for $\Gamma^\#$, where $D$ is a suitable initial guess for $\Gamma^\#$. We choose
\begin{equation}\label{eq:InitialGuess}
D_{ij}=\frac{1}{w^{{\rm acc.}}_i} \delta_{ij} = \Delta t_i \delta_{ij}
\end{equation}
which ensures the (linear) convergence of the series for the considered class of Markov processes\cite{suppmaterialAIP}. We can therefore truncate the series at some finite $M$ (the truncation limit) and arrive at
\begin{equation}\label{eq:TACexpansion}
\begin{split}
 c_R&\approx 2(v, D \sum\limits_{n=0}^M(1 - \Gamma D)^nu)\\ &= 2  \sum\limits_{ij}
 \delta R_i \Delta t_i \left(\sum\limits_{n=0}^M(P_{\rm kMC})^n\right)_{ij} \delta R_j  P_{j,{\rm stat.}}.
 \end{split}
\end{equation}
where we have used $(1 - \Gamma D)_{ij}=P_{\rm kMC}(i|j)$. Generally, a good choice of $M$ is unknown, and needs to be determined by testing the convergence. 

As $(P_{\rm kMC})^n_{ij}$ is the probability to be in state $i$ after $n$ kMC steps starting in state $j$, the TAC can be written as an expected value of a discrete time Markov chain, with the initial distribution $P_{i_0,{\rm stat.}}$
\begin{equation}\label{eq:TACMarkovChain}
\begin{split}
  c_R\approx & 2  \sum\limits_{i_0, i_1 \ldots i_M} \Bigg[   \left(\sum\limits_{n=0}^M \delta R_{i_n} \Delta t_{i_n} \delta R_{i_0}\right)  \\
  & \times P_{\rm kMC}(i_M|i_{M-1}) \ldots P_{\rm kMC}(i_1|i_0)P_{i_0,{\rm stat.}} \Bigg].
\end{split}
\end{equation}
We could thus run a kMC trajectory to obtain samples $i_0$ with their corresponding weights ($\propto \Delta t_{i_0}$) and simulating a single discrete time Markov chain for each initial $i_0$. Fortunately, the later is not necessary as the next $M$ states in a kMC trajectory are generated according to $P_{\rm kMC}$. We can save the time and just employ these. Putting this together with Eq. \ref{eq:StatkMCAv}, we arrive at
\begin{equation}\label{eq.TACEstimate}
 c_R\approx  \frac{2}{T} \sum\limits_{l=M}^{N} \delta R_{i_l} \Delta t_l\sum\limits_{n=0}^{M}  \delta R_{i_{l-n}} \Delta t_{l-n}
\end{equation}
where $T=\sum\limits_{l=0}^{N-M} \Delta t_l$. 

An equation as \ref{eq.TACEstimate} could also be motivated by first representing $\delta R(t)$ by a suitable trajectory averaging and then performing the (truncated) time integration. Choosing $M$ large enough and a subsequent time step averaging results in the same formulas as Eq. \ref{eq:TACMarkovChain}. This allows to interpret $M$ as a measure of the autocorrelation time in number of kMC steps. Multiplying it with the average time step then leads to a physical autocorrelation time. 

When implementing Eq. \ref{eq.TACEstimate}, the need to sum over the last $M$ steps could become the computational burden, when $M$ becomes large. A straight forward idea to lift this would be to employ a circular array which always stores $\delta R \Delta t$ of the last $M$ steps and to introduce an variable $S$ for the sum, which is updated in every step $l$ according to the rule $S \rightarrow S - \delta R_{i_{l-M}} \Delta t_{l-M} + \delta R_{i_{l}} \Delta t_{l}$. Then the element in the circular array containing $\delta R_{i_{l-M}} \Delta t_{l-M}$ is overwritten by $\delta R_{i_{l}} \Delta t_{l}$. This allows an update in $\mathcal{O}(1)$ CPU-time and has $\mathcal{O}(M)$ storage requirement. However, for stiff problems as the CO oxidation on RuO$_2$(110), the time steps $\Delta t_{l}$ easily vary by some orders of magnitude (and $M$ can become large). We then end up in adding and subtracting small numbers from the variable $S$, which might quickly cause inaccuracies due to the finite machine precision. We therefore complement
the circular array with a binary tree. The leaves then carry the elements of the circular array and an interior node stores the sum of the values stored at its  children. The root node then carries  the desired sum over all leaves. Choosing $M$ as a power of two allows to employ a so-called perfect binary tree, i.e. every interior node has exactly two children and all leaves have the same depth\cite{sane2007data}. This data structure and the corresponding operations can efficiently be implemented in a single array, which is of more relevance in kMC simulations than maximum flexibility in choosing $M$. If we now update the leave carrying $\delta R_{i_{l-M}} \Delta t_{l-M}$, only the ancestor nodes need to be updated which scales $\mathcal{O}(\log_2 M)$. Being a little less efficient than the straight forward circular array, we, however, have the advantage, that at no point we need to subtract two numbers. The supplementary material provides pseudocode for sampling of the TAC during the kMC simulation\cite{suppmaterialAIP}.

\subsection{Linear Response Theory based direct sampling of the DRS}\label{sec:LinearResponse}
The second step is the direct sampling of the sensitivities using our estimator based on Linear Response Theory. 
Linear Response Theory deals with small, in general time-dependent perturbations of the generator $\Gamma$ in the master equation\cite{hanggi1982stochastic}. For our purposes, the perturbations result from  changing the rate constants, i.e. we multiply the RC  $k^\alpha$ with a factor $(1+ \varepsilon^\alpha(t))$. The perturbed generator is then given by
\begin{equation}
\Gamma_R(t)=\Gamma + \Gamma^\alpha \varepsilon^\alpha(t).
\end{equation}
As we are interested in the stationary processes, we can restrict to the case, when the system was initially in the steady state for $\varepsilon^\alpha(t)=0$, i.e. $P_j(t=0)=P_{j,{\rm stat.}}$.
In the linear response regime, when $\varepsilon^\alpha(t)$ is small, the time dependent deviation  $\delta \langle R \rangle (t)$ from the stationary expectation obeys\cite{hanggi1982stochastic}
\begin{equation}\label{eq:LRF1}
\begin{split}
\langle R \rangle (t) - \langle R \rangle_{\rm stat.}= &\int\limits_0^t  \chi^\alpha (t-s) \varepsilon^\alpha(s)ds
\end{split}
\end{equation}
where we omitted an explicit dependence of $R$ on $\varepsilon^\alpha(t)$ as we target at an estimate for $X^\alpha_1$.
The Linear Response Function (LRF) $ \chi^\alpha (t)$ in Eq. \ref{eq:LRF1} is given by
\begin{equation}\label{eq:defSusz}
\begin{split}
\chi^\alpha(t) &= \sum\limits_{ij}\delta R_i(t) \Gamma^\alpha_{ij}P_{j,{\rm stat.}} \\ &=  \sum\limits_{ijk}\delta R_i (e^{\Gamma t})_{ij}\Gamma^\alpha_{ik}P_{k,{\rm stat. }}.
\end{split}
\end{equation}
and can be obtained from the properties of the stationary initial state and the unperturbed process. If  $\varepsilon^\alpha(t)=const.$ for $t>0$, we expect that $\langle R \rangle (t)$ converges
to a new steady state and, as we are in the linear regime with respect to the perturbation $\varepsilon^\alpha$, we have
\begin{equation}\label{eq:DefIRF}
 X^{\alpha}_1=\frac{1}{\langle R \rangle} \int\limits_0^\infty  \chi^\alpha (s)ds=\left ( \tilde R, \Gamma^\#  \Gamma^\alpha P_{\rm stat.}\right)
\end{equation}
where $\tilde R_i=\delta R_i/\langle R \rangle$ and $\Gamma^\#$ is the same generalized inverse as in Eq. \ref{eq:TACPseudoInverse}, because we have the same orthogonality properties
for $\Gamma^\alpha P_{\rm stat.}$ as for $\delta R_i  P_{i,{\rm stat.}}$ (compare section \ref{sub:estimatorTAC} and the supporting information\cite{suppmaterialAIP}).

\subsubsection{Estimator for $ X^{\alpha}_1$}\label{sub:estimatorIRF}
As in section \ref{sub:estimatorTAC}, we expand $\Gamma^\#$ into a series and arrive at
\begin{equation}
 X^\alpha_1 \approx \bar X^\alpha_1= \left(\tilde R , D \sum\limits_{n=0}^M(P_{\rm kMC})^n  \Gamma^\alpha P_{\rm stat.}\right),
\end{equation}
with $D$ as defined by Eq. \ref{eq:InitialGuess}. We employ the same series expansion in above equation as for the estimation of the TAC in section \ref{sub:estimatorTAC}. The truncation limit $M$ should therefore be the same for both and we can employ a reasonable choice obtained in the first step also here. 

Next, we decompose $ \Gamma^\alpha$ into its diagonal part $\Gamma^{\alpha,D}_{ij}=-\delta_{ij}  \sum\limits_l w^\alpha_{lj}$ and the off-diagonal
partial transition matrix and make the decomposition $\bar X^\alpha_1 = X^\alpha_D + X^\alpha_W$, where
\begin{equation}\label{eq:XW}
\begin{split}
 X^\alpha_W& \approx \sum\limits_{i_{-1},i_0, \ldots i_M} \Bigg[  \left(\sum\limits_{n=0}^M \tilde R_{i_n} \Delta t_{i_n} O^\alpha_{i_0,{i_{-1}}}\right)\\
 & \times P_{\rm kMC}(i_M|i_{M-1}) \ldots P_{\rm kMC}(i_0|i_{-1})P_{i_{-1},{\rm stat.}}\bigg].
 \end{split}
\end{equation}

Here, we have have used that $P_{\rm kMC}(i|j)=0 \Rightarrow w^\alpha_{ij}=0$ and introduced the rescaled partial transition matrix $O^\alpha$ defined by
\begin{equation}
\begin{split}
 &w^\alpha_{ij}= P_{\rm kMC}(i|j)O^\alpha_{ij},\text{ i.e }~  O^\alpha_{ij}= \begin{cases}
                 w^{{\rm acc.}}_j &\text{if} ~ j\rightarrow i \in \alpha\\
                 0 & \text{else}
                \end{cases} .
\end{split}
\end{equation}
Note,  that in Eq. \ref{eq:XW} $X^\alpha_W$ is an expected  value over a discrete Markov chain of length $M+2$ instead of $M+1$ as in Eq. \ref{eq:TACMarkovChain}.
The contribution from the diagonal part of $\Gamma^\alpha$ can be written as
 \begin{equation}\label{eq:XD}
 \begin{split}
  X^\alpha_D & \approx \sum\limits_{i_{-1}i_0,\ldots i_M} \Bigg[ \left(\sum\limits_{n=-1}^{M} \tilde R_{i_n} \Delta t_{i_n} w^{\alpha,{\rm acc.}}_{i_{-1}}\right) \\
  & \times P_{\rm kMC}(i_M|i_{M-1}) \ldots P_{\rm kMC}(i_0|i_{-1})P_{i_{-1},{\rm stat.}}\Bigg]
  \end{split}
 \end{equation}

where we have added a term $\left(\tilde R , D (1 + \Gamma D)^{M+1}  \Gamma^{\alpha,D} P_{\rm stat.}\right)$, so that the Markov chains in Eqn. \ref{eq:XW} and \ref{eq:XD} have the same length. We can do so as this term converges to zero for large enough $M$\cite{suppmaterialAIP}.

Putting equations \ref{eq:XW} and \ref{eq:XD} together and employing the same argumentation as in section \ref{sub:estimatorTAC}, we arrive at the estimator
\begin{equation}\label{eq:IRFEstimator}
\begin{split}
 X^\alpha_1 & \approx \frac{1}{T} \sum\limits_{n=M+1}^{N}\Bigg[ - \tilde R_{i_{n}} w^{\alpha,{\rm acc.}}_{i_{n}}\Delta t_n^2 \\
 &+ \tilde R_{i_{n}}\Delta t_n\sum\limits_{l=1}^{M+1} (O^\alpha_{i_{n-l+1}i_{n-l}}- w^{\alpha,{\rm acc.}}_{i_{n-l}}) \Delta t_{n-l} \Bigg],
 \end{split}
\end{equation}
where $T=\sum\limits_{l=0}^{N-M-1} \Delta t_l$. We term this way to estimate $X^\alpha_1$ Integrated Response Function (IRF) approach. Although being based on a very different derivation, the IRF estimator \ref{eq:IRFEstimator} is very similar to the estimator presented in ref. \cite{Cao633827,Cao701341}, except that it employs deterministic $\Delta t$ instead of random time steps $\delta t$, which might help to improve the performance as for log-likelihood estimators\cite{jcp_1.3690092,jcp1.4905957}. However, the motivation from generalized inverses and series expansion might open ways to improve its performance, e.g. by choosing a different initial guess $D$.

The above formula \ref{eq:IRFEstimator} explains why we have added the extra term in Eq. \ref{eq:XD}: Both $O^\alpha_{i_{n+1}i_{n}}$ and $w^{\alpha,{\rm acc.}}_{i_{n}}$ can now be calculated at the kMC step $n$. The  accumulated rate needs to be calculated anyways and for $O^\alpha_{i_{n+1}i_{n}}$ we just need to know which type of reaction will be executed next. We can therefore employ the same binary tree for summation as for the TAC sampling with negligible overhead. The leaves will now carry the values $(O^\alpha_{i_{n+1}i_{n}} -w^{\alpha,{\rm acc.}}_{i_{n}})\Delta t_{n} $ instead of $\delta R_{i_{n}} \Delta t_{n}$.

Pseudocode for sampling of the IRF can be found in the supplementary material\cite{suppmaterialAIP}.

\subsection{Coupled Finite Differences}\label{sec:CoupledFiniteDifferences}
The direct sampling of the sensitivities, is not always successful. For some DRS, we might still have to large sampling errors.
We estimate these DRSs, which are not sufficiently accurate, using Finite Difference Approximations (FDA). In particular we employ central Finite Differences
\begin{equation}
 \left(\frac{\partial \langle R \rangle (\{k^\beta\})}{\partial  \log{k^\alpha}} \right)_{k^{\beta\neq \alpha}}\approx \frac{ \langle R \rangle ( (1+ h){k^\alpha}  ) - \langle R \rangle ( (1- h){k^\alpha} )}{2h}
\end{equation}
which are second order accurate in the  difference parameter $h$. Straightforward FDA uses two independent kMC simulations to estimate $\langle R \rangle (\ldots (1+ h){k^\alpha}\ldots  )$ and  $\langle R \rangle (\ldots (1- h){k^\alpha}\ldots  )$. As $h$ approaches zero, the difference between  $\langle R \rangle (\ldots (1+ h){k^\alpha}\ldots  )$ and  $\langle R \rangle (\ldots (1- h){k^\alpha}\ldots  )$ will become smaller than the sampling errors in the statistically independent estimates, and the FDA might thereby carry a huge error. To overcome this to a certain degree, the two simulation could be coupled such that the fluctuations of the two estimates point into the same direction and the estimated difference has a smaller sampling error than the two estimates for $\langle R \rangle$. The simplest of such couplings is the Common Random Number approach\cite{Rathinam10p034103}, where   we would employ the same set of pseudo random numbers for both kMC simulations. 
This approach is easy to implement, but seems
not to provide significant improvements for lattice systems\cite{Arampatzis2014}. More advanced coupling strategies exist (see e.g. \cite{Arampatzis2014}), of which we choose the Coupled Finite Differences (CFD) proposed by Anderson\cite{Anderson12p2237}, for the ease of its implementation.

To outline how CFD can be applied to lattice problems, it is convenient to decompose the set $\beta$ into different reaction channels. That is, two transitions $i_1 \rightarrow j_1$ and $i_2 \rightarrow j_2$ belong to the same reaction channel $A_\beta$, if they belong to the same reaction $\beta$ and cause the same change $\xi^{A_\beta}:= j_1-i_1=j_2 -i_2$. Intuitively, this introduces a spatial resolution of the reaction set $\beta$. While $\beta$ includes, for instance, all CO adsorption on cus sites, a CO adsorption on the first cus site will not belong to the same reaction channel as a CO adsorption on the second cus site, simply because both change different entries in the integer vector representing the current state of the system. In contrast, all transitions, which correspond to a  CO adsorption on the first cus site, will be in the same channel as they all change the same entry from {\em empty} to {\em CO} and leave all others unchanged.
With this concept, we can rewrite the master equation as
\begin{equation}\label{eq:MEPropensities}
 \frac{d}{dt} P_i(t) = \sum\limits_\beta \sum\limits_{A_\beta} \left[ a^{A_\beta}(i-\xi^{A_\beta})P_{i -\xi^{A_\beta}}(t) - a^{A_\beta}(i)P_{i }(t)\right]
\end{equation}
where, in the context outlined in section \ref{sec:General},  the reaction propensity $a^{A_\beta}(i)$ has the form
\begin{equation}\label{eq:Propensity}
 a^{A_\beta}(i)=\begin{cases}
        k^\beta, ~  \text{if } ~i \rightarrow i +\xi^{A_\beta} \in A_\beta\\
       0, \quad \text{else}
       \end{cases}
\end{equation}
The two independent trajectories for an uncoupled Finite Difference Approximation can now be be regarded as a single process operating on two lattices. The corresponding master equation is then
\begin{equation}\label{eq:MEPropUncoupled}
\begin{split}
\frac{d}{dt} P_{i,j}(t) &= \sum\limits_\beta \sum\limits_{A_\beta}\left[ a_h^{A_\beta}(i-\xi^{A_\beta})P_{i -\xi^{A_\beta},j}(t) -  a_h^{A_\beta}(i)P_{i,j }(t)\right]\\
			&+  \sum\limits_\beta \sum\limits_{A_\beta}\left[  a_{-h}^{A_\beta}(j-\xi^{A_\beta})P_{i ,j-\xi^{A_\beta}}(t) -  a_{-h}^{A_\beta}(j)P_{i,j }(t)\right]
\end{split}
\end{equation}
where $P_{i,j}(t)$ is the joint probability to find the first lattice in state $i$ and the second lattice in state $j$. The index $h$ at the propensity $a_{h}^{A_\beta}(\cdot)$ indicates that this is obtained by using the RCs $\{ \ldots,(1+ h){k^\alpha} ,\ldots\}$. Here $k^\alpha$ is the RC, for which we want to estimate the corresponding DRS. Summing over $j$ will therefore lead to the master equation \ref{eq:MEPropensities} for the RCs $\{ \ldots,(1+ h){k^\alpha} ,\ldots\}$ and summing over $i$ will achieve the same for $\{ \ldots,(1- h){k^\alpha} ,\ldots\}$. Correspondingly, $\langle R \rangle ( (1+ h){k^\alpha}  )$ will be estimated by using only the $i$-component of the joint process and $\langle R \rangle ( (1- h){k^\alpha}  )$ and by using only the $j$-component.
If we now calculate the variance of the difference $R_{h}(i) - R_{-h}(j)$ using the joint probability $P_{i,j}$ we obtain
\begin{equation}
 {\rm Var}( R_{h}-R_{-h})={\rm Var}(R_h) +{\rm Var}(R_{-h} )- 2\langle \delta R_{h}\delta R_{-h}\rangle,
\end{equation}
i.e. the variance of the difference would be reduced if there would be a positive correlation $\langle \delta R_{h}\delta R_{-h}\rangle$ between $R_{h}(i)$ and $ R_{-h}(j)$. For uncoupled FDA obeying Eq. \ref{eq:MEPropUncoupled} the correlation $\langle \delta R_{h}\delta R_{-h}\rangle$ is, of course,  zero due to statistical independence of $i$ and $j$. The idea is now to introduce a new joint process, which gives the same averages $\langle R_{h} \rangle$ and $\langle R_{-h} \rangle$ but a positive correlation  $\langle \delta R_{h}\delta R_{-h}\rangle$. There are multiple different possibilities to realize this\cite{Arampatzis2014}. Anderson's method employs a process obeying the master equation
\begin{equation}\label{eq:ME_CFD}
 \begin{alignedat}{2}
 \frac{d}{dt} P_{i,j}(t) &=& \sum\limits_\beta \sum\limits_{A_\beta}\left[ \right. & a_1^{A_\beta}(i-\xi^{A_\beta}, j-\xi^{A_\beta})P_{i -\xi^{A_\beta},j-\xi^{A_\beta}}(t) \\
			& 	&~					       & \left. -  a_1^{A_\beta}(i,j)P_{i,j }(t)\right]\\
			& + & \sum\limits_\beta \sum\limits_{A_\beta}\left[ \right. & a_{2}^{A_\beta}(i-\xi^{A_\beta},j)P_{i -\xi^{A_\beta},j}(t)  \\
			& 	&					       & \left. -  a_{2}^{A_\beta}(i,j)P_{i,j }(t)\right]\\
			& + &\sum\limits_\beta \sum\limits_{A_\beta}\left[ \right. & a_3^{A_\beta}(i, j-\xi^{A_\beta})P_{i,j-\xi^{A_\beta}}(t)  \\
			& 	&~					       & \left. -  a_3^{A_\beta}(i,j)P_{i,j }(t)\right]
 \end{alignedat}
\end{equation}

with the propensities
\begin{equation}\label{eq:Propensities_CFD}
\begin{split}
 a_1^{A_\beta}(i,j) & =  \min( a_h^{A_\beta}(i),a_{-h}^{A_\beta}(j))\\
 &=\begin{cases}
                                                                 k_{-h}^\beta\quad  &\begin{split}&\text{ if }  i\rightarrow i+\xi^{A_\beta} \in  A_\beta \\ &\text{ and } j\rightarrow j+\xi^{A_\beta}\in  A_\beta \end{split}\\
                                                                 ~\\
                                                                 0 &\text{ else}
                                                                \end{cases}
                                                            \\~\\
 a_{2}^{A_\beta}(i,j)&= a_h^{A_\beta}(i) - \min( a_h^{A_\beta}(i),a_{-h}^{A_\beta}(j))\\
 &=\begin{cases}
                                                                 k_{h}^\beta  \quad &\begin{split}& \text{ if }  i\rightarrow i+\xi^{A_\beta} \in  A_\beta\\ & \text{ and } j\rightarrow j+\xi^{A_\beta} \notin  A_\beta \end{split}\\
                                                                  ~\\
                                                                 k_{h}^\beta - k_{-h}^\beta   &\begin{split}&\text{ if }  i\rightarrow i+\xi^{A_\beta} \in  A_\beta \\ & \text{ and } j\rightarrow j+\xi^{A_\beta}\in  A_\beta \end{split}\\
                                                                  ~\\
                                                                 0 &\text{ else}
                                                                \end{cases}
\\~\\
a_{3}^{A_\beta}(i,j) &= a_{-h}^{A_\beta}(i) - \min( a_h^{A_\beta}(i),a_{-h}^{A_\beta}(j))\\
&=\begin{cases}
                                                                 k_{-h}^\beta \quad &\begin{split}& \text{ if }  i\rightarrow i+\xi^{A_\beta} \notin  A_\beta \\ & \text{ and } j\rightarrow j+\xi^{A_\beta} \in  A_\beta \end{split}\\
                                                                 0 &\text{ else}
                                                                \end{cases}
\end{split}
\end{equation}
where we inserted Eq. \ref{eq:Propensity} for the respective second equal sign. Using that, we can split the propensity $a_{2}^{A_\beta}(i,j)$ into two new ones $a_{2,1}^{A_\beta}(i,j)$ and $a_{2,2}^{A_\beta}(i,j)$. The first will be non-zero ($a_{2,1}^{A_\beta}(i,j)=k_{h}^\beta$) only when the first case fulfilled ($ i\rightarrow i+\xi^{A_\beta}\in  A_\beta$ and $j\rightarrow j+\xi^{A_\beta}\notin  A_\beta$). The second will only be non-zero ($a_{2,2}^{A_\beta}(i,j)=k_{h}^\beta - k_{-h}^\beta$) for the second case ($ i\rightarrow i+\xi^{A_\beta} \in  A_\beta$ and $j\rightarrow j+\xi^{A_\beta}\in  A_\beta$).

We can describe the Markov process defined by Eqn. \ref{eq:ME_CFD} and \ref{eq:Propensities_CFD} in a more chemical language, which allows an implementation using the standard interfaces of lattice kMC codes.
If we have the sites $(a,b,c,\ldots)$ for our original problem, the process for CFD operates on a lattices with the sites $(a_1,b_1,c_1\ldots,a_2,b_2,c_2\ldots)$. For the CO oxidation on RuO$_2$(110), this can be achieved by replacing the two site types $(\rm br,cus)$ by the four sites  $(\rm br_1,cus_1, br_2,cus_2)$. Suppose the reaction channel $A_\beta$ originally corresponds to a change of the adsorption states on the sites $(a,b)$, i.e. it can be represented by
\begin{equation}\label{eq:OrgProcess}
 {\rm A}@a +{\rm B}@b  \overset{k^\beta}\longrightarrow {\rm C}@a + {\rm D}@b
\end{equation}
where  ${\rm A}@a$ means that the site $a$ carries the adsorbate ${\rm A}$, which is altered to the adsorbate ${\rm C}$ after the process has been executed.
On our doubled lattice for CFD, we have to replace this with appropriate new reaction channels.
The reaction channel corresponding to $a_1^{A_\beta}$ executes the original process \ref{eq:OrgProcess} simultaneously on both lattices, i.e. we have to introduce the reaction
\begin{equation}\label{eq:CFDProcess1}
\begin{split}
 &{\rm A}@a_1 +{\rm B}@b_1 + {\rm A}@a_2  +{\rm B}@b_2 \\
                         & \overset{k_{-h}^\beta~}  \longrightarrow  {\rm C}@a_1 + {\rm D}@b_1  + {\rm C}@a_2 + {\rm D}@b_2 ~,
\end{split}
\end{equation}
where $\overset{k_{-h}^\beta~}\longrightarrow $ indicates that this will be executed with a rate constant $k_{-h}^\beta$.
Correspondingly, the reaction channel for $a_{2,2}^{A_\beta}$ will have the reaction equation
\begin{equation}\label{eq:CFDProcess2}
\begin{split}
 & {\rm A}@a_1 +{\rm B}@b_1 + {\rm A}@a_2  +{\rm B}@b_2 \\
 &\overset{k_{h}^\beta - k_{-h}^\beta }\longrightarrow {\rm C}@a_1 + {\rm D}@b_1 + {\rm A}@a_2  +{\rm B}@b_2~,
  \end{split}
 \end{equation}
 as it changes only the configuration on the first lattice, but requires this change to be possible on the second lattice.
 What remains are the channels, which correspond to $a_{2,1}^{A_\beta}$ and $a_{3}^{A_\beta}$. These change the configuration on one lattice according to $A_\beta$ and require that $A_\beta$ is not possible on the respective other lattice. The required logic is usually not implemented in standard interfaces of lattice kMC codes (at least, {\sc kmos} does not allow this). We therefore employ the same trick which lead to splitting of $a_{2}^{A_\beta}$ and introduce a new reaction channel for each situation which fulfills the requirements. In other words, we introduce for every pair $(E,F)\neq(A,B)$ the reaction
 \begin{equation}\label{eq:CFDProcess3}
  \begin{split}
 & {\rm A}@a_1 +{\rm B}@b_1 + {\rm E}@a_2  +{\rm F}@b_2 \\
 &\overset{k_{h}^\beta}\longrightarrow {\rm C}@a_1 + {\rm D}@b_1 + {\rm E}@a_2  +{\rm F}@b_2~,
  \end{split}
  \end{equation}
  which corresponds to $a_{2,1}^{A_\beta}$ and the reaction
  \begin{equation}\label{eq:CFDProcess4}
  \begin{split}
 & {\rm E}@a_1 +{\rm F}@b_1 + {\rm A}@a_2  +{\rm B}@b_2 \\
 &\overset{k_{-h}^\beta~}\longrightarrow {\rm E}@a_1 + {\rm F}@b_1 + {\rm C}@a_2  +{\rm D}@b_2~,
  \end{split}
 \end{equation}
which corresponds to $a_{3}^{A_\beta}$.
We can thus employ the standard interface of {\sc kmos} for defining the process and perform the CFD without needing to touch the source code. The price we pay is that for a two site reaction as \ref{eq:OrgProcess} in the original process, we have to introduce $\mathcal{O}((N_{\rm spec.})^2)$ in the process for the CFD. Here, $N_{\rm spec.}$ is the number of surface species or more general the number of states a site can attain. Hence, this is probably not the most efficient way of implementing CFD and more complex reactions operating on more than two sites will require even more additional reactions. However, for the case at hand, this overhead stays reasonable and running $N$ kMC steps for the CFD comes roughly at five times the cost of running $N$ kMC steps of the original process.

\section{Results}\label{sec:Results}

As described in section \ref{sub:COoxidation}, we study the CO oxidation on the RuO$_2$(110) surface for the reaction conditions $T$=600\,K, $p_{\text{O}_2}=1$\,bar and  $p_{\text{CO}} \in [0.05,50]$\,bar.
The investigated reactions rate is the CO oxidation turnover frequency (displayed in Figure \ref{TOFfig}) and we test its sensitivity to all 22 different rate constants using the outlined three stage strategy. 

\subsection{First step: bounds}\label{sub:TACResults}
\begin{figure}
\includegraphics[width=\linewidth]{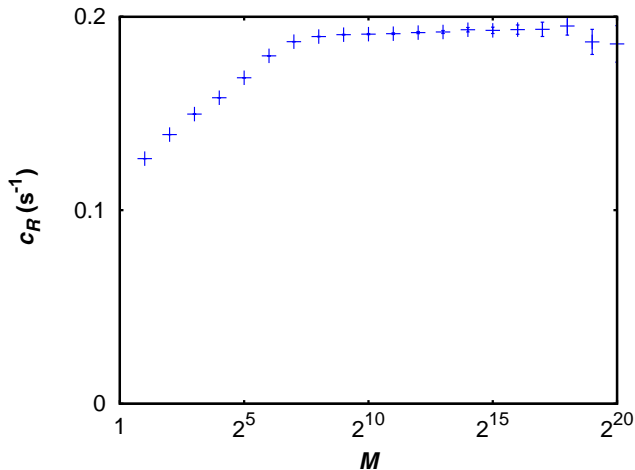}
\caption{Estimate of time-integrated auto-correlation function (TAC) as a function of the expansion parameter $M$ for the CO oxidation on RuO$_2$(110) for  $T$=600\,K, $p_{\text{O}_2}=1$\,bar and  $p_{\text{CO}}=1$\,bar.  Also shown is the standard deviation as error bars. The TAC is well converged for a choice $M=2^{12}=4096$.
\label{Correlation_vs_M}}
\end{figure}
The first step is the estimation of bounds for the DRS from the Fisher Information Metric (FIM) and the time-integrated autocorrelation function (TAC).
In all simulations, we employ a lattices of $20\times 20$ unit cells as in previous studies \cite{reuter2006first,meskine2009examination}. After an initial relaxation to steady state with $10^8$ steps, the TAC is sampled from 100 sub-trajectories each having a length of $10^7$ steps and using $10^7$ decorrelation steps  between the sampling trajectories.

In practice, a good choice for $M$ is usually unknown and thus we have to test different $M$ for estimating the TAC $c_R$ until we reach convergence. As the choice of $M$ does not alter the kMC code, we can perform
this test for a range of possible values during a single kMC simulation. Although we do not expect small values of $M$ to be sufficient, we test all powers of two between $2^0=1$ and $2^{20}\approx 10^6$.
For the case $p_{\text{CO}}=1$\,bar, we show the dependence of the TAC estimates on the choice of $M$ in Figure \ref{Correlation_vs_M}. Also shown is the standard deviations of the TAC estimates as error bars. For small values of $M$, these are very small but increase for larger values of $M$. Thus the choice of $M$ has to balance two errors. It needs to be large enough such that expansion \ref{eq:TACexpansion} is accurate, and, on the other hand, small enough such that the variance of the sampled estimate does not grow too large. This can be seen in Figure \ref{Correlation_vs_M}, where the TAC is well converged for a choice $M=2^{12}=4096$, and the higher expansion accuracy for larger $M$ gets corrupted by the sampling error.

\begin{figure}
\includegraphics[width=\linewidth]{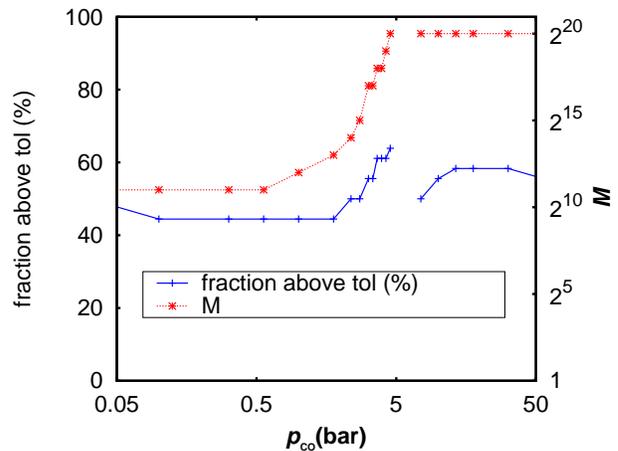}
\caption{Fraction of all reactions with a  sensitivity bound $B^\alpha$ above $2\%$ (blue, left axis) and the corresponding value for the expansion parameter $M$ (red, right axis).
\label{DRS_varPCO_bound}}
\end{figure}

All DRSs must add up to one and we therefore expect the relevant DRSs to be in this order of magnitude\cite{suppmaterialAIP}. As a threshold for the relevance, we choose $tol=0.02$, i.e. all DRSs with a bound $B^\alpha$ smaller than this will not be calculated later on. Figure \ref{DRS_varPCO_bound} shows the fraction of all DRSs, which bounds $B^\alpha$ are above the threshold $tol$ (blue,left axis) in dependence of $p_{\text{CO}}$ and for fixed $p_{\text{O}_2}=1$\,bar and  $T$=600\,K. Further, we indicate our choice for the truncation limit $M$ (red, right axis). Notably, we found no convergence for $M \leq 2^{20}$ for CO partial pressures around $5$\,bar, i.e in the vicinity of the point of highest reactivity. The interpretation of the expansion \ref{eq:TACexpansion} as a  time integration suggests that we are close to a second order phase transition, where the correlation time diverges and thus the integration limit must be chosen very large. A closer look at the coverage curves in ref. \cite{temel2007does} corroborate this interpretation, but also the results on spatial correlation from refs. \cite{herschlag2015,Gelss2016} point into this direction. Also, we can work with much smaller truncation limits $M$ for small $p_{\text{CO}}$ than for large, although, in both cases, we are not close to the reaction conditions for which we expect a second order phase transition. We explain this as follows. For low $p_{\text{CO}}$, in the O-covered regime, the kinetics takes place almost exclusively on the cus sites\cite{meskine2009examination} and the surface lattice is fully covered with oxygen. An adsorption of CO onto a vacant site resulting from an oxygen desorption  triggers with very high probability either an CO desorption or a CO$+$O reaction in the next step, as the RCs for both are much higher than for any other possible reactions. The resulting vacancies are likely to be filled with oxygen (due to the high O$_2$ adsorption RC), we  thus return quickly to a fully oxygen covered surface. We therefore expect the correlation time in number of kMC steps to be rather short. For high $p_{\text{CO}}$ and the concomitant almost always fully CO-covered surface, the removal of oxygen would require much longer as now there are many fast competing processes. For instance, CO desorption from cus sites has a ten times higher RC than the most likely $  {\rm O_{cus} + CO_{br} } $ reaction and there are many CO$_{\rm cus}$. Thus the decay of the oxygen adsorbed state would take rather many kMC steps.

In all cases, we found that roughly $50\%$ of the DRS bounds are above our tolerance. For the case, that we would directly proceed with a finite difference approximation as in ref. \cite{10.1371/journal.pone.0130825}, this would translate into a $50\%$ saving in CPU-time. However, we perform a direct sampling in between and the question naturally arises what is the benefit of running an extra kMC simulation to save $50\%$ of the sampling in another. The answer is that this step is not only there for ruling out DRSs but also to identify a reasonable choice of $M$ as, in the next section, we will employ the same expansion which led to Eq. \ref{eq:TACexpansion}. Thus, this pre-screening step reduces the sampling effort to $50\%/N_M$ as we save the examination of $N_M$ different values of $M$. As we need to sample only a single TAC but $N_{reac.}$ DRSs, determining $M$ in this step allows for significant savings in the  subsequent step.

\subsection{Second step: direct sampling of the DRS}\label{sub:IRFResults}
\begin{figure}
\includegraphics[width=\linewidth]{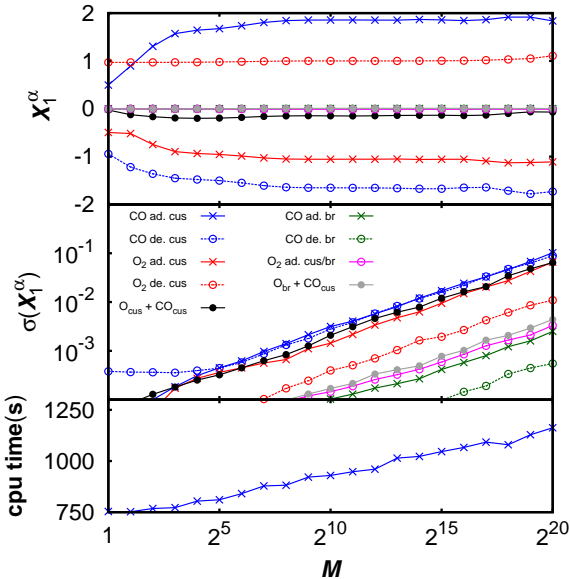}
\caption{
    Estimates of the Degree of Rate Sensitivities (DRS) based on linear response sampling as a function of the expansion parameter $M$ for the CO oxidation on RuO$_2$(110) for  $T$=600\,K, $p_{\text{O}_2}=1$\,bar and $p_{\text{CO}}=1$\,bar (top panel). As the TAC (see Fig \ref{Correlation_vs_M}), the DRSs are converged for a choice $M=2^{12}=4096$. The standard deviations of the sampled DRSs grow with the root of $M$ (middle panel). The lower panel shows the CPU time, which grows, as expected, linear with the logarithm of $M$ (up to an additive constant).
\label{IRF_vs_M}}
\end{figure}

As in section \ref{sub:TACResults}, we consider first the case $T$=600\,K, $p_{\text{O}_2}=1$\,bar and $p_{\text{CO}}=1$\,bar and the dependence on the expansion length $M$. For this, we ignore the results from the bound estimations and sample all DRSs. In order to test the computational costs, we, however, perform an extra kMC simulation for each $M$. For these,  we employ the similar settings as in section \ref{sub:TACResults}, i.e. a $20 \times 20$ lattice and sampling over 100 snippets of $10^7$ kMC steps, between which we perform $10^6$ steps for decorrelation. As in section \ref{sub:TACResults}, we employ these 100 samples to obtain the estimated DRSs and corresponding variances. The results are summarized in Figure \ref{IRF_vs_M}. As this is the demanding task, we only show the $X^\alpha_1$ in the top panel. The estimates are well converged for $M=2^{12}$, i.e. the value we have extracted from the simulation of the TAC. The variance of the estimation increases linearly with $M$, which is reflected by the standard deviation, shown in the panel below. For the current set of reaction conditions and a reasonable choice of $M$,
the standard deviations are in the order of $10^{-2}$ or below, so that we estimate all DRSs for the cost of two extra kMC simulations (one for estimating the bounds and $M$ and one for estimating the DRSs).
The CPU-time for the last simulation comes at only slightly higher costs by the extra sampling as can be seen in the lowest panel in Fig. \ref{IRF_vs_M}. As expected, the sampling overhead depends linearly on $\log{M}$ and, for $M=2^{20}\approx 10^{6}$, it requires roughly one third of the CPU time.  The computational savings of the two-steps procedure stem from the possibility to avoid the test of different values of $M$. Without this prescreening of different $M$ values, the sampling overhead would be responsible for $>80\%$ of the CPU load for the considered case. The simulation would therefore run roughly six times as long as a pure kMC simulation. Utilizing the estimated bounds obtained in section \ref{sub:TACResults}, the overhead could have been further reduced by roughly $50\%$. The sampling of the TAC comes at roughly the same costs as a full DRS sampling at fixed $M$, as we have to test for multiple $M$. So, we end up with an overhead of a factor $2-3$ (including the CPU-time for the prescreening). Without prescreening we instead would arrive of an overhead of a factor $\approx 6$. For more complicated reaction mechanism and the concomitant larger number of RCs, this saving would increase.

\begin{figure}
\includegraphics[width=\linewidth]{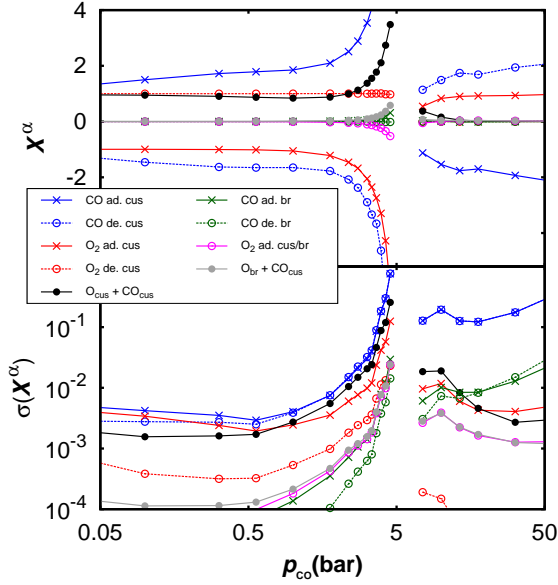}
\caption{
The sampled Degree of Rate Sensitivity (DRS) based on linear response sampling as function of the CO partial pressure $p_{\text{CO}}$ ($T$=600\,K, $p_{\text{O}_2}=1$\,bar, top panel). The fluctuations during sampling expressed as standard deviation $\sigma$ are shown in the lower panel. The simulation employed the expansion parameters $M$ obtained in the previous section (compare Fig. \ref{DRS_varPCO_bound}). Not shown are those DRS which are significantly below zero and the range around $p_{\text{CO}}=5$\,bar, for which $M>2^{20}$ would be required.
\label{IRF_vs_pCO}}
\end{figure}

With the estimated bounds and the values for $M$ obtained in section \ref{sub:TACResults}, we now sample the DRSs for varying CO partial pressure. We estimate these from $1000$ sub-trajectories each of the size $100\times M$ steps, but at least $10^6$ and at most $10^7$. As $M$ can be regarded as a correlation time, we employ $10\times M$ decorrelation steps between the samples. As for Fig. \ref{DRS_varPCO_bound}, we do not perform an estimation around the point of highest activity, since we would need $M$ to be larger than one million. In principle, larger $M$ could have been employed, but, as the variance increases linearly with $M$, we do not expect accurate enough estimates.

The results for the DRSs are shown in the top panel of Fig. \ref{IRF_vs_pCO}. Shown are only those DRSs, which are non-zero in the range of the considered reaction conditions. In most cases, the number of non-vanishing DRSs is small, even considering the pre-screening. Inspecting the bottom panel of Fig. \ref{IRF_vs_pCO}, we find that the standard deviation of our estimates depends largely on the considered reaction. Especially, those DRSs, which are low, can be estimated pretty well with very small sampling errors. However, also most other DRSs can be obtained with an accuracy below $0.1$ ,i.e. reasonably accurate for many purposes. Only close to the gap and for the CO ad/desorption on cus sites at high CO partial pressures, the standard deviations exceed the value of $0.1$. This increase in the standard deviations corresponds directly to the larger values of $M$ required there. For high $p_{\text{CO}}$, we observe that the DRS estimates for the fastest processes (CO ad/desorption on cus sites) have the highest variance. This behaviour is similar to the Girsanov transform based approach\cite{jcp1.4905957}. If we need more accurate estimates with $\sigma(X^\alpha_1) < 0.1$, we could increase the sampling time. However, the sampling error behaves as $\mathcal{O}(N^{-\frac{1}{2}})$, where $N$ is the total number of kMC steps. So pushing the error below the targeted limit might become very demanding for those DRSs, which can not be sampled well, and it might become reasonable to employ a different strategy for obtaining these.

\subsection{Third step: Coupled Finite Differences}
\begin{figure}
\includegraphics[width=\linewidth]{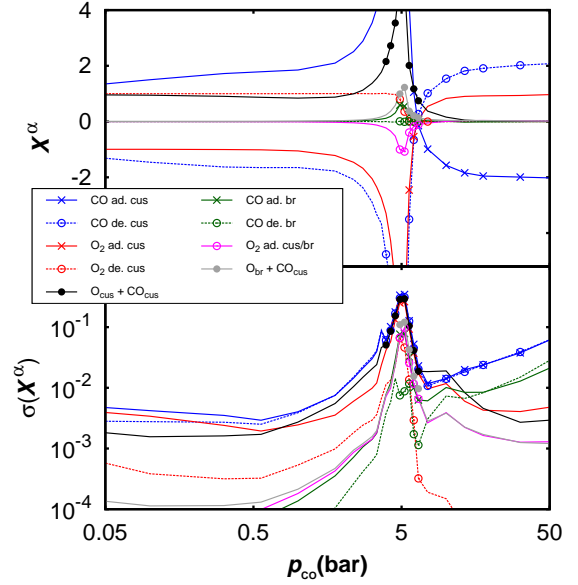}
\caption{ Same as Fig. \ref{IRF_vs_pCO}, but missing points (around $p_{\text{CO}}=5$\,bar) and those with a too high variance have been determined using Coupled Finite Differences. This data is represented using symbols, all other points have been obtained from linear response sampling.
\label{all_vs_pCO}}
\end{figure}

We now employ the CFD to estimate the DRSs, which could not be sampled with sufficient accuracy by the direct sampling approach from the previous section. These are all DRSs which lay in the gap around $p_{\text{CO}}=5$\,bar and those for which the existing estimates have a standard deviation above $0.1$. To estimate the DRSs and the respective variances, we employ $1280$ samples each having $10^6$ kMC steps. All other settings are as in section \ref{sub:IRFResults}. For the difference parameter, we choose $h=0.01$.

Figure \ref{all_vs_pCO} shows the final results combining the IRF and CFD estimates (top panel) with the corresponding standard deviation (lower panel). Symbols mark data points which have been obtained using CFD, all others have been directly sampled by IRF. The standard deviations for the estimates are now below the desired value of $0.1$, except close to the point of highest activity where for some DRSs we stay a little above the desired accuracy. However, the affected DRSs are rather large there and the relative error is of an acceptable order.

Having a closer look, we find that for the most of the pressure range in Fig. \ref{all_vs_pCO} we just need a few or no CFD estimates. This holds especially true for low $p_{\text{CO}}$ where all DRSs can accurately be estimated and no CFD needs to be performed. This can be explained by the relatively small values for the expansion order $M$ in the IRF, as the variance of the estimator grows linear with $M$ and thus small $M$ should rather lead to accurate estimates. But also for large $p_{\text{CO}}$ only the DRSs for CO ad/desorption on cus sites needed improvements. Only close to the pressures where the DRSs for CO and O$_2$ ad/desorption on cus sites ``diverge'' multiple DRSs needed refinements using CFD.

The procedure, as outlined, leaves room for improvement if CPU time is really the limiting factor. Most notably, we could perform the IRF with much less kMC steps. At low $p_{\text{CO}}$, the DRSs have an accuracy which is roughly ten times lower than required, so we could save a factor around hundred here. Also, we would not need to perform ten billion kMC steps for the IRF sampling at high CO partial pressures. Except for CO ad/desorption on cus sites, all DRSs would have the desired accuracy already at around one billion steps. Also running IRF with $M>10^6$ in the gap around $p_{\text{CO}}=5$\,bar, might not help a lot for the large DRSs but it might allow to determine the lower ones with sufficient accuracy.
However even in this unoptimized form, we significantly reduce the computational effort with respect to the very expensive numerical derivatives.

\section{Discussion}
\begin{figure}
\includegraphics[width=\linewidth]{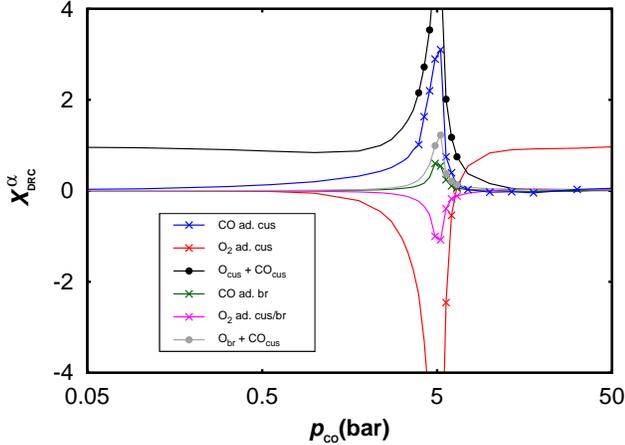}
\caption{
The sampled Degree of Rate Control as function of the CO partial pressure $p_{\text{CO}}$ ($T$=600\,K, $p_{\text{O}_2}=1$\,bar, top panel). Symbols represent data which was obtained using Coupled Finite Differences.
\label{campbell_vs_pCO}}
\end{figure}

We now turn to the discussion of the results obtained in the previous sections. As it is the more common measure, we calculated the  Degree of Rate Control (DRC) from our DRS results and Eq.  \ref{eq:WhyWeShouldAlwaysCalculateOurDRCs}. The DRC is shown in Fig. \ref{campbell_vs_pCO}. For low CO partial pressures, the DRC identifies only a single rate-determining step, which is the $\rm CO_{cus} + O_{cus}$ reaction. For high CO partial pressures, the oxygen adsorption on cus-sites is the only rate-determining step. Between these two extremes, we find multiple steps, which have significant DRCs. Close to $5$\,bar, the absolute DRCs of all displayed reactions are in the order of one or larger and the notion of a single rate-determining step seems not appropriate, anymore. 

\begin{figure}
\includegraphics[width=\linewidth]{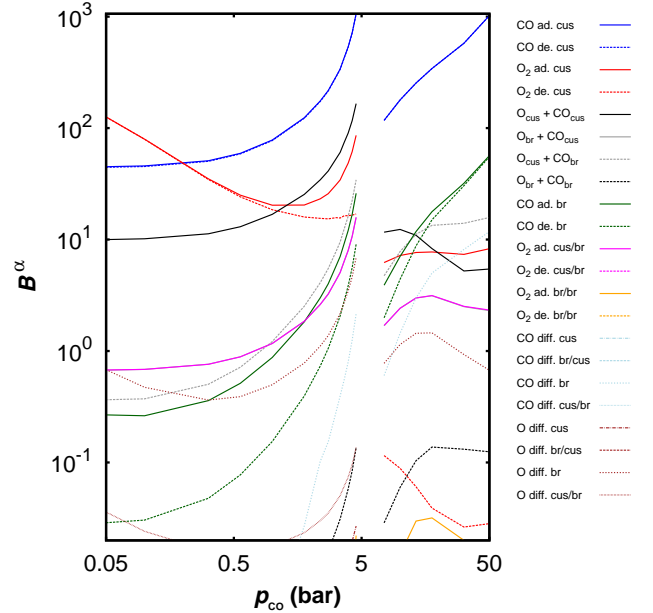}
\caption{
Bounds $B^{\alpha}$ for the Degree of Rate Sensitivity  as function of the CO partial pressure $p_{\text{CO}}$ ($T$=600\,K, $p_{\text{O}_2}=1$\,bar, top panel).  The bounds show the same rapid increase close to $p_{\text{CO}}=5$\,bar as the DRS. However, some elementary steps would be marked important by the bounds, but are not according to the DRS criterion, e.g. adsorption and desorption of $\rm CO$ on bridge sites. 
\label{all_bounds_pCO}}
\end{figure}

If we compare the DRC in Fig. \ref{campbell_vs_pCO} with the DRS in Fig. \ref{all_vs_pCO}, the DRC provides a clearer picture. In contrast, the DRS plot is more busy, but also provides information which is not present in the DRCs. At low $p_{\text{CO}}$, the DRS identifies adsorption and desorption of CO and oxygen on cus-sites to have an impact on the TOF, with opposite signs but (roughly) the same absolute values for forward and backward reaction. This allows for a intuitive microscopic interpretation. At these conditions, the surface is fully oxygen covered and we need to get CO onto the surface for a high TOF with the employed Langmuir-Hinshelwood mechanism. To achieve this, empty sites need to be created by one of the oxygen desorption processes and the most effective $2\text{O}_{\text{cus}} \rightarrow \text{O}_2(\text{gas})$ desorption has a positive DRS. Now, CO and oxygen adsorption onto the created empty sites compete with each other. So, the $\text{O}_2(\text{gas}) \rightarrow 2\text{O}_{\text{cus}}$ adsorption has a negative DRS as it blocks the more catalytically active configuration. The $\text{CO}$ adsorption on cus sites has a positive sign, because it leads to configurations for which the $\text{O}_{\text{cus}} + \text{CO}_{\text{cus}} \rightarrow \text{CO}_2(\text{gas})$ reaction is possible. Once a $\text{CO}$ molecule is adsorbed on a cus site, we have a competition between desorption and the actual reaction.
The desorption destroys the desired configuration and therefore has a negative sign. The $\text{O}_{\text{cus}} + \text{CO}_{\text{cus}} \rightarrow \text{CO}_2(\text{gas})$ has a positive sign, which now results from the two competing parts $X^\alpha_0$ and $X^\alpha_1$. The direct relative contribution $X^\alpha_0$ to the TOF is always positive. At these low CO partial pressures the  $\text{O}_{\text{cus}} + \text{CO}_{\text{cus}} \rightarrow \text{CO}_2(\text{gas})$ reaction is responsible for almost all the reactivity and its relative contribution is therefore very close to one. This can be seen from Fig. \ref{all_bounds_pCO}, which shows the sensitivity bounds $B^\alpha$ (Eq. \ref{eq:bounds}), which are proportional to the square roots of reaction rates of each elementary step.  The second is the contribution $X^\alpha_1$ to the creation/annihilation of the desired configurations, which should be negative for this case because the reaction removes the CO from the surface. This contribution is close to zero as the dominant channel of the CO removal from the surface is the CO desorption from cus sites, which has a higher frequency (compare Fig. \ref{all_bounds_pCO}). 

At low CO partial pressures, the cancellation of the DRS for ad/desorption on cus sites suggests that these steps are close to equilibrated. This interpretation is supported by the bounds for the forward and backward rates, which are almost the same for CO ad/desorption and oxygen ad/desorption on cus sites, respectively. As we increase $p_{\text{CO}}$, the CO coverage on cus-sites also increases. Although this coverage is still very low, the fast $\text{O}_{\text{cus}} + \text{CO}_{\text{cus}} \rightarrow \text{CO}_2(\text{gas})$ reaction becomes an alternative to the $\text{O}_{\text{cus}}$-desorption for oxygen removal from the surface. This is reflected by the deviation of the bounds for $\text{O}_{\text{cus}}$-desorption and $\text{O}_{\text{cus}}$-adsorption in Fig. \ref{all_bounds_pCO}. In consequence, the DRC for the pair of $\text{O}_{\text{cus}}$-ad/desorption starts deviating from zero, as they are not equilibrated anymore. For the point at $p_{\text{CO}}=5$\,bar, we speculated that its very close to a second order phase transition and the correlation time as well as fluctuations should increase as we approach this point. This results in an increase of the time-integrated auto-correlation (TAC), which is reflected in an increase of the bounds for most elementary steps. The same holds for most integrated response functions $X^\alpha_1$, and, in consequence, the absolute DRS for the relevant elementary steps increase while approaching the ``critical point''. For the DRC in Fig. \ref{all_bounds_pCO} for the $\text{CO}_{\text{cus}}$-ad/desorption, this results in a amplification of the effect of the slight deviation from equilibrium. The $\text{O}_{\text{cus}}$-ad/desorption is much better equilibrated at low $p_{\text{CO}}$ and therefore this amplification of the DRC sets in at little higher $p_{\text{CO}}$. In contrast to the $\text{CO}_{\text{cus}}$-ad/desorption, the increase in the absolute DRC of the $\text{O}_{\text{cus}}$-ad/desorption is more pronounced, because this reaction pair is additionally strongly driven out of equilibrium while approaching $p_{\text{CO}}=5$\,bar. Notably, the DRC for the $\text{O}_{\text{cus}} + \text{CO}_{\text{cus}} $ reaction shows a similar increase, when the CO partial pressure exceeds the oxygen partial pressure $p_{\text{O}_2}=1$\,bar by a factor two to three. This means that the contribution $X^\alpha_1$ by creation and annihilation of desired configuration must be positive for the $\text{O}_{\text{cus}} + \text{CO}_{\text{cus}} $ reaction. 
This is in contrast to the case at low partial pressures, where it is negative (but close to zero). At  higher pressures ($p_{\text{CO}} \in[2,5]$\,bar), the presence of CO on the surface is still critical as reflected by the positive DRS for the $\text{CO}_{\text{cus}}$-adsorption. However, the removal of $\text{CO}_{\text{cus}}$ by the $\text{O}_{\text{cus}} + \text{CO}_{\text{cus}} $ reaction is now compensated by the creation of two vacant cus-sites, which are are likely to be filled by $\text{CO}$.

Close to $p_{\text{CO}}=5$\,bar, oxygen still dominates the surface and getting CO onto the surface is still crucial, but we have already some kind of mixed composition with both
$\text{CO}$ and oxygen on the surface. We now have some CO on the bridge sites, and consequently the fast $\text{O}_{\text{cus}} + \text{CO}_{\text{br}} $ reaction starts contributing
to the overall TOF (compare the supplementary material \cite{suppmaterialAIP}). Oxygen is very strongly bound to the bridge sites and therefore covers these. The CO adsorption on
bridge site has a positive DRS for two reasons: i) it enables the $\text{O}_{\text{cus}} + \text{CO}_{\text{br}} $ reaction and ii) it prevents oxygen to adsorb on the pair of vacant
sites created by the later reaction. The corresponding $\text{CO}_{\text{br}} $-desorption plays no role, because the $\text{O}_{\text{cus}} + \text{CO}_{\text{br}} $ reaction is significantly faster and
removes most of the $\text{CO}_{\text{br}}$. The $\text{O}_{\text{cus}} + \text{O}_{\text{br}}$ adsorption removes vacant sites, which are needed for the $\text{CO}$-adsorption 
on both types of sites, resulting in a negative DRS. The $\text{O}_{\text{br}} + \text{CO}_{\text{cus}} $ reaction has two effects. First it contributes slightly to the overall TOF, second it it creates free bridge and cus sites, which are both likely to be filled with CO. The $\text{O}_{\text{cus}} + \text{CO}_{\text{br}} $ reaction itself has no impact, because it is so fast that no other mechanism, which would destroy the $\text{O}_{\text{cus}}$/$\text{CO}_{\text{br}} $ pair, can compete with it.

Increasing the CO partial pressure beyond $5$\,bar, the DRS and DRC of all species drop, because we are moving away from the point with the highest correlation and large fluctuations. Further, we are increasing the CO content on the surface, so that it becomes less and less important to get CO onto the surface. At around $6-7$\,bar, we reach a point, where all DRS and DRC are close to zero except for the $\text{O}_{\text{cus}} + \text{CO}_{\text{cus}} $ reaction. As this is still the dominant contribution to the overall TOF, we can interpret this point as the best mixed phase, where the appearance of appreciable configurations  is largely unaffected  by accelerating one of the elementary steps. A further increase leads then to a CO dominated surface, where the DRC identifies the $\text{O}_{\text{cus}}$-adsorption as the rate-determining step. Having a closer look at the results for the DRS, this result is easily understood. On this CO covered surface, the crucial point is to get oxygen onto the surface. Therefore the $\text{CO}_{\text{cus}}$-desorption has a positive impact, as it creates empty sites. Oxygen adsorption onto cus-sites competes with CO adsorption onto the same sites, i.e. the oxygen adsorption has  a positive DRS and the CO adsorption has a negative DRS. At high CO partial pressures, CO ad/desorption is equilibrated and their DRS must cancel when summing to the DRC. Once on the surface, oxygen binds very strongly and is almost exclusively removed by forming CO$_2$, with dominant reaction being the $\text{O}_{\text{cus}} + \text{CO}_{\text{br}} $ and smaller contributions by the $\text{O}_{\text{cus}} + \text{CO}_{\text{cus}} $ reaction. Due to the stronger binding of CO to bridge sites (compared to cus-sites), $\text{CO}_{\text{br}}$-desorption is comparatively slow, so that vacancy pairs with one bridge site are rare and thus the competition between oxygen adsorption involving a bridge site and the $\text{CO}_{\text{br}}$-adsorption does not play a big role.

Finally, we want to draw our attention to the question, which information could be extracted from only considering the bounds in Fig. \ref{all_bounds_pCO} as sensitivity measures.   
As expected, the bounds overestimate the DRS, often by more than one order of magnitude. On the plus side, the five most important steps are correctly identified for low CO pressures. Also the steep increase of the most absolute DRS is properly reflected by the bounds. The O$_\text{cus}$ desorption does not follow this trend and this is properly reproduced by the corresponding DRS bound in Fig. \ref{all_bounds_pCO}. For high CO pressures, the adsorption and desorption of CO on cus-sites are ranked most important, but the oxygen adsorption on cus-sites is ranked only on place seven according to the Fisher Information Metric (FIM) based criterion. Furthermore, a number of processes involving bridge sites are identified as important. We would therefore miss the important feature that only the kinetics on the cus sites determines the reaction rate for  most of the considered reaction conditions. So, a FIM based sensitivity analysis\cite{Pantazis2013} provides a first, more global insight into the reaction kinetics, but a detailed picture of the microscopic reaction pathways requires a deeper analysis.

\section{Conclusion}
We have presented a three stage procedure to determine local sensitivity indices, specifically the Degree of Rate Sensitivity, from which other sensitivity measures as the Degree of Rate Control can easily be derived. In the first stage, bounds for the DRSs are obtained to eliminate those DRSs which are close to zero. In the second stage, the remaining DRSs are estimated using a direct sampling approach, which is based on an truncated series expansion of time-integrated Linear Response Functions. Only those DRSs which can not accurately enough be  estimated in the previous stages, are finally obtained using Coupled Finite Differences. The devised approach can lead to significant computational savings, compared with the straightforward approach employing solely numerical derivatives.

We have demonstrated the approach on the example of the CO oxidation on RuO$_2$(110). Comparing with with our previous study for the sensitivity analysis of the same model\cite{meskine2009examination}, we arrive at significantly more accurate estimates for the sensitivity indices in a fraction of the CPU-time. For this model, we have also demonstrated, that the detailed information obtained during the three stage procedure can be used to derive a detailed interpretation of the nature of the microscopic reaction kinetics. We found that the DRS criterion is a suitable complement to the more common DRC and allows to explain, why a step is rate-determining. 

The methods for sampling the bounds and the IRF approach have been implemented into the open-source kMC package {\sc kmos} and we are currently preparing it to become part of the next publicly available update. The CFD has been implemented using just the front-end and future development will target at allowing for a more efficient implementation but also at alternative coupling strategies\cite{Arampatzis2014}. Also some effort  needs to be directed towards a more balanced strategy minimizing the overall computational costs.

\section*{Supplementary Material}
See supplementary material for additional information on the derivations of series expansions of the TAC and the IRF and the pseudocode for their sampling as well as for the DRS sum rule.

\begin{acknowledgments}
This research was carried out in the framework of {\sc Matheon} supported by Einstein Foundation Berlin.
\end{acknowledgments}

%

\pagebreak
\widetext
\begin{center}
\textbf{\large Supplementary material to A practical approach to the sensitivity analysis for kinetic Monte Carlo simulation of heterogeneous catalysis}
\end{center}
\setcounter{equation}{0}
\setcounter{figure}{0}
\setcounter{table}{0}
\setcounter{page}{1}
\makeatletter
\renewcommand{\theequation}{S\arabic{equation}}
\renewcommand{\thefigure}{S\arabic{figure}}
\renewcommand{\bibnumfmt}[1]{[S#1]}
\renewcommand{\citenumfont}[1]{S#1}

\section{Preliminaries}\label{sec:Preliminaries}
Throughout this supplementary material, we denote (column) vectors with bold  and matrices with large letters. Elements of both are indicated by subscripts. 

We restrict ourselves to Markov processes over a discrete, finite state space with $N+1$ states. We assume, that the generator $ \Gamma_{ij}=w_{ij} -w^{{\rm acc.}}_j\delta_{ij} $ has only a single stationary distribution $\pmb P_{\rm stat.}$ and
\begin{equation}\label{eq:RightEV_ME}
 \lim\limits_{t\rightarrow \infty} (e^{\Gamma t})_{ij}=P_{{\rm stat.},i}=:(\pmb{b}_0)_i
\end{equation}
which is fulfilled for irreducible, ergodic processes\cite{kelly1979}.
Further, we require that the kMC transition probability $P_{{\rm kMC},ij}= w_{ij}/w^{{\rm acc.}}_j$ obeys the similar law and repeated application converges to a unique
(discrete time) probability distribution $\pmb h$, i.e.
\begin{equation}\label{eq:PkMC_primitive}
 \lim\limits_{n\rightarrow \infty}(P_{{\rm kMC}}^n)_{ij}= (\pmb h)_i
\end{equation}
which holds if $ P_{{\rm kMC}}$ defines an aperiodic discrete-time Markov chain\cite{meyer2000matrix}. 

$\pmb{b}_0$ is a right eigenvector of the generator $\Gamma$ to the eigenvalue zero and $h$ is a right eigenvector of $P_{{\rm kMC}}$ for the eigenvalue one.
The corresponding left eigenvector is in both cases
\begin{equation}\label{eq:LeftEV}
\pmb  b'_0=(1,1,\ldots,1)^T .
\end{equation}
From the definition of $\Gamma$ and $P_{{\rm kMC}}$ it follows that $\pmb{b}_0$ and $\pmb h$ obey the relation
\begin{equation}\label{eq:h_vs_b}
 D\pmb h= a \pmb b_0
\end{equation}
with the diagonal matrix $D_{ij}= 1/w^{{\rm acc.}}_j \delta_{ij}$ and a real number $a$.
We introduce the projections $S=\pmb{b}_0 \pmb{b}_0'^T$ and $Q=1-S$, where $1$ is to be understood as the identity matrix. 
These have the properties
\begin{equation}\label{eq:PropProjections}
 \Gamma S= S \Gamma=0, \quad \Gamma Q= Q \Gamma=\Gamma.
\end{equation}

\section{Series expansion for $\int\limits_0^\infty (\pmb v,e^{\Gamma t} \pmb u)dt$}
We now want to prove that for the processes in section \ref{sec:Preliminaries}, the identity 
\begin{equation}
 \begin{split}
 \lim\limits_{t \rightarrow \infty} \int\limits_0^t(\pmb{v}, e^{\Gamma s}   \pmb{u}) ds=-(\pmb{v}, \Gamma^\# \pmb{u})=\lim\limits_{n \rightarrow \infty} (\pmb{v}, D \sum\limits_{l=0}^n(1 - \Gamma D)^l \pmb{u})
 \end{split}
\end{equation}
holds for $(\pmb{v},\pmb{b}_0)=0$ and $(\pmb{b}_0,\pmb{u})=0$, where $(\pmb a, \pmb b)=\pmb a^T\pmb b$ is the standard scalar product. This is the case for $v_i=R_i - \langle R \rangle$ and $u_i=(R_i - \langle R \rangle) P_{{\rm stat.},i}$, which leads to the time-integrated auto-correlation (TAC), or for $v_i=R_i - \langle R \rangle$ and $u_i=(\Gamma^\alpha P_{{\rm stat.}})_i$, which leads to the integrated linear response (IRF). Here $\langle \cdot \rangle$ is the stationary expectation (i.e. using $ P_{{\rm stat.}}$), $R$ is an arbitrary observable, and $\Gamma^\alpha$ is a partial generator defined in section (1) of the main manuscript. The generalized inverse $\Gamma^\#$ will be defined below.
Since $\pmb  v \bot \pmb b_0$ and $\pmb u \bot \pmb b'_0$, we have the identity
\begin{equation}
 (\pmb{v}, e^{\Gamma s}   \pmb{u})= (Q^T \pmb{v},e^{\Gamma s}   Q \pmb{u}).
\end{equation}

The generator $\Gamma$ is not invertible as one eigenvalue is zero. We therefore introduce the bases
\begin{equation}
 \begin{array}{c}
  B= (\pmb{b}_0,\pmb{b}_1, \ldots \pmb{b}_N)\\
  B'=(\pmb{b}'_0,\pmb{b}'_1, \ldots \pmb{b}'_N)
 \end{array}
 ~\text{with } ~ (\pmb{b}_i,\pmb{b}_j)=\delta_{ij},
\end{equation}
where $\pmb{b}_0$ and $\pmb{b}'_0$ are right and the left eigenvectors of $\Gamma$ (as given by eqn. \ref{eq:RightEV_ME} and \ref{eq:LeftEV}).
We now represent $\pmb{v}$ in the basis $B'$ and $\pmb{u}$ in the basis $B$, i.e. $\pmb{v}=\sum\limits_{k=0}^N \tilde v_k \pmb{b}'_k$ and $\pmb{u}=\sum\limits_{k=0}^N \tilde u_k \pmb{b}_k$. Exploiting that $\sum\limits_{k=0}^N \pmb{b}_k \pmb{b}'^T_k=1$, we find
\begin{equation}
 (\pmb{v}, e^{\Gamma s}   \pmb{u})=(\tilde {\pmb{v}},\tilde Q e^{\tilde\Gamma s} \tilde Q \tilde {  \pmb{u}})
\end{equation}
where $\tilde\Gamma_{ij}=(\pmb{b}'_i,\Gamma \pmb{b}_j)$ and $\tilde Q_{ij}=  (\pmb{b}'_i,Q \pmb{b}_j)$, i.e. we have the representation
\begin{equation}
 {\LARGE\mbox{{$\tilde\Gamma$}}}= \left(
\begin{array}{cc}
  0 & ~~0~~ \cdots~ ~0 \\ 
  0 & \raisebox{-15pt}{{\LARGE\mbox{{$A$}}}} \\[-2ex]
  \vdots & \\
  0 &
\end{array}
\right), \quad {\LARGE\mbox{{$e^{\tilde\Gamma t}$}}}= \left(
\begin{array}{cc}
  1  & ~~0~~ \cdots~ ~0 \\  
  0 & \raisebox{-15pt}{{\LARGE\mbox{{$e^{At}$}}}} \\[-2ex]
  \vdots & \\
  0 &
\end{array}
\right).
\end{equation}
As we implied only a single stationary state the $N\times N$ submatrix $A$ is now invertible. $\tilde Q$ is a diagonal matrix with a zero on the first diagonal element and all other diagonal elements being one. We get for the integral
\begin{equation}\label{eq:IntExp}
 {\LARGE\mbox{{$\int\limits_0^t \tilde Qe^{\tilde \Gamma s} \tilde Qds$}}}= \left(
\begin{array}{cc}
  0  & ~~0~~ \cdots~ ~0 \\  
  0 & \raisebox{-15pt}{{\large\mbox{{$\int\limits_0^t e^{As}ds$}}}} \\[-2ex]
  \vdots & \\
  0 &
\end{array}
\right)=\left(
\begin{array}{cc}
  0  & ~~0~~ \cdots~ ~0 \\  
  0 & \raisebox{-15pt}{{\large\mbox{{$A^{-1}e^{At} - A^{-1}$}}}} \\[-2ex]
  \vdots & \\
  0 &
\end{array}
\right).
\end{equation}
By the requirement \ref{eq:RightEV_ME}, $e^{\Gamma t}$ converges to the stationary distribution for $t\rightarrow \infty$ and $e^{At}$ must therefore converge to zero. Thus
\begin{equation}\label{eq:IntLimit}
 {\LARGE\mbox{{$\lim\limits_{t\rightarrow\infty} \int\limits_0^t \tilde Qe^{\tilde \Gamma s} \tilde Qds$}}}= \left(
\begin{array}{cc}
  0  & ~~0~~ \cdots~ ~0 \\  
  0 & \raisebox{-15pt}{{\large\mbox{{$ - A^{-1}$}}}} \\[-2ex]
  \vdots & \\
  0 &
\end{array}
\right)=: {\LARGE\mbox{{$-\tilde\Gamma^\#$}}}.
\end{equation}
which defines the generalized inverse $\Gamma^\#$ by backtransformation to our standard basis. 
We now employ the Neumann series for $A^{-1}$
\begin{equation}\label{eq:VonNeumann}
 -A^{-1}= A^{-1}_0 \sum\limits_{n=0}^\infty (1_N +AA^{-1}_0)^n
\end{equation}
where $1_N$ is the $N\times N$ unit matrix and  $A^{-1}_0$ is a suitable initial guess. The Neumann series converges (linearly), if $1_n +AA^{-1}_0$ has spectral radius smaller than one.
Now, working in this $N\times N$ setting does not help as we do not know $b_0$ nor can we compute any linear algebra standard operation for the targeted problems as $n$ is usually way too large. We therefore extend to $N+1$ dimensions and eliminate the dependence on the bases $B$ and $B'$. We rewrite the generalized inverse
\begin{equation}
\begin{split}
  {\LARGE\mbox{{$-\tilde\Gamma^\#$}}}&=\left(
\begin{array}{cc}
  0  & ~~0~~ \cdots~ ~0 \\  
  0 & \raisebox{-15pt}{{\large\mbox{{$A^{-1}_0 \sum\limits_{n=0}^\infty (1_N +AA^{-1}_0)^n$}}}} \\[-2ex]
  \vdots & \\
  0 &
\end{array}
\right)=\\ 
&=
\left(\begin{array}{cc}
  0  & ~~0~~ \cdots~ ~0 \\  
  0 & \raisebox{-15pt}{{\large\mbox{{$A^{-1}_0 $}}}} \\[-2ex]
  \vdots & \\
  0 &
\end{array}\right)
 \left(
\begin{array}{cc}
  0  & ~~0~~ \cdots~ ~0 \\  
  0 & \raisebox{-15pt}{{\large\mbox{{ $\sum\limits_{n=0}^\infty (1_N +AA^{-1}_0)^n$}}}} \\[-2ex]
  \vdots & \\
  0 &
\end{array}
 \right)=\\
 &=
{\LARGE\mbox{{$\tilde Q \tilde \Gamma^{-1}_0\tilde Q \sum\limits_{n=0}^\infty$ }}}\left(
\begin{array}{cc}
  0  & ~~0~~ \cdots~ ~0 \\  
  0 & \raisebox{-15pt}{{\large\mbox{{ $ (1_N +AA^{-1}_0)$}}}} \\[-2ex]
  \vdots & \\
  0 &
\end{array}
\right)^n\\
&={\LARGE\mbox{{$\tilde Q \tilde \Gamma^{-1}_0\tilde Q \sum\limits_{n=0}^\infty \left[\tilde Q ( 1 + \tilde \Gamma \tilde Q \tilde \Gamma^{-1}_0\tilde Q)\tilde Q \right]^n$}}}
\end{split}
\end{equation}
where we have used that multiplying with $Q$ from both sides leads to zeros in the first row and the first column for arbitrary $(N+1)\times (N+1)$ matrices $G$, i.e. we get a matrix with $A^{-1}_0$ in the right place by $\tilde Q \tilde \Gamma^{-1}_0\tilde Q$ for a suitably chosen $\Gamma^{-1}_0$. Transforming back to the standard basis we get
\begin{equation}\label{eq:Series1}
\begin{split}
 \lim\limits_{t \rightarrow \infty} \int\limits_0^t(\pmb{v}, e^{\Gamma s}   \pmb{u}) ds&= \left(\tilde {\pmb{v}}, \tilde Q \tilde \Gamma^{-1}_0\tilde Q \sum\limits_{n=0}^\infty \left[\tilde Q ( 1 + \tilde \Gamma \tilde Q \tilde \Gamma^{-1}_0\tilde Q)\tilde Q \right]^n \tilde {\pmb{u}}\right)\\
 &=\left( {\pmb{v}},  Q  \Gamma^{-1}_0 Q \sum\limits_{n=0}^\infty \left[ Q ( 1 +  \Gamma Q  \Gamma^{-1}_0 Q) Q \right]^n  {\pmb{u}}\right)\\
 &=\left( {\pmb{v}},  Q  \Gamma^{-1}_0 Q \sum\limits_{n=0}^\infty \left[ Q ( 1 +  \Gamma   \Gamma^{-1}_0 )^n Q \right] {\pmb{u}}\right),
 \end{split}
\end{equation}
where we have used that $Q^n=Q$ and $Q\Gamma=\Gamma=\Gamma Q$ for the last equality.

We now choose $\Gamma_0^{-1}$ such that we can approximate the infinite series in \ref{eq:Series1} with a finite sum. In other words, the series must be convergent. For this, we set $\Gamma_0^{-1}=D$, where $D$ is the diagonal matrix with $D_{ii}=1/w_i^{acc}$ defined in section \ref{sec:Preliminaries}. With this choice, the series  in \ref{eq:Series1} converges if \cite{meyer2000matrix}
\begin{equation}\label{eq:convergence1}
 \lim\limits_{n \rightarrow \infty} (Q(1 +\Gamma D )Q)^n =  \lim\limits_{n \rightarrow \infty} QP_{\rm kMC}^nQ= \lim\limits_{n \rightarrow \infty}( QP_{\rm kMC}^n - QP_{\rm kMC}^n \pmb{b}_0 \pmb{b'}_0^T)= 0
\end{equation}
By the requirement \ref{eq:PkMC_primitive}, the limit $\sum\limits_{n=0}^\infty P_{\rm kMC}^n = \pmb{h} \pmb{b'}_0^T$  exists. The limit in eq. \ref{eq:convergence1} then becomes
\begin{equation}\label{eq:convergence2}
\lim\limits_{n \rightarrow \infty}( QP_{\rm kMC}^n - QP_{\rm kMC}^n \pmb{b}_0 {\pmb{b}'}_0^T)= \lim\limits_{n \rightarrow \infty}( QP_{\rm kMC}^n) - \lim\limits_{n \rightarrow \infty} QP_{\rm kMC}^n \pmb{b}_0 {\pmb{b}'}_0^T= Q\pmb{h} {\pmb{b}'}_0^T - Q\pmb{h} {\pmb{b}'}_0^T\pmb{b}_0 {\pmb{b}'}_0^T=0
\end{equation}
where we used $(\pmb{b}'_0, \pmb{b}_0)=1$ and which proves the convergence. We can therefore approximate 
\begin{equation}
 \int\limits_0^\infty (\pmb{v},e^{\Gamma t} \pmb{u})dt \approx (\pmb{v}, Q D Q\sum\limits_{n=0}^M Q(1 -\Gamma  D )^n Q \pmb{u})
\end{equation}
for sufficiently large $M$.

We now eliminate $Q$ from $\left( {\pmb{v}},  Q  D Q \sum\limits_{n=0}^\infty \left[ Q ( 1 +  \Gamma  D )^n Q \right] {\pmb{u}}\right)$. For this, we employ  $Q^T\pmb{v}=\pmb{v}$ and $Q\pmb{u}=\pmb{u}$ to arrive at
\begin{equation}\label{eq:removeLastQ}
 (\pmb v, Q D Q(1 +\Gamma  D )^nQ \pmb u)= (\pmb v,D Q (1 +\Gamma D)^n \pmb u).
\end{equation}
For the last $Q$, we show by induction  that ($S=1-Q$)
\begin{equation}\label{eq:RgivesZero}
 S (1 +\Gamma D)^n \pmb u=0,
\end{equation}
which implies that we can replace $Q$ by $1$ in equation \ref{eq:removeLastQ}.
The base case $n=0$ holds, as $S\pmb u=\pmb{b}_0 {\pmb{b}'}_0^T\pmb u=0$ by the requirement $ ({\pmb{b}'}_0, \pmb u)=0$. For the induction step (hypothesis $S(1-\Gamma D)^{n}\pmb u=0$), we employ that $S(1-\Gamma D)=S$ as $S\Gamma=0$ (eq. \ref{eq:PropProjections}). We then have $S(1-\Gamma D)^{n+1}\pmb u=S(1-\Gamma D)(1-\Gamma D)^{n}\pmb u=S(1-\Gamma D)^{n}\pmb u$, which is zero by the induction hypothesis.

Putting all together, we get the following approximate expression 
\begin{equation}\label{eq:QEliminated}
\lim\limits_{t \rightarrow \infty} \int\limits_0^t(\pmb{v}, e^{\Gamma s}   \pmb{u}) ds \approx \left( \pmb v,  \sum\limits_{n=0}^M D(1- \Gamma D)^n \pmb u \right) .
\end{equation}
which does not require the knowledge of $b_0$ and we can evaluate integrated correlation functions or linear response function using our standard basis. Note that this holds irrespective of the choice of $D$, as long this leads to a convergent series in eq. \ref{eq:Series1}. 

\section{$\lim\limits_{M \rightarrow \infty} \left(\tilde{ \pmb{R}} , D (1 + \Gamma D)^{M}  \Gamma^{\alpha,D} \pmb{P}_{\rm stat.}\right)=0$}
In the linear response section of the main manuscript, we have used that the terms 
\[\left(\tilde{ \pmb{R}} , D (1 + \Gamma D)^{M}  \Gamma^{\alpha,D} \pmb{P}_{\rm stat.}\right)\] tend to zero for large $M$. This allowed us to approximate the contributions from the diagonal part $\Gamma^{\alpha,D}$ of the perturbation $\Gamma^\alpha$ and its off-diagonal part by Markov chains of the same length. 
As $( \pmb{b}'_0,\Gamma^{\alpha,D} \pmb{P}_{\rm stat.})$ is not necessary zero, we consider this limit separately. Using eq. \ref{eq:PkMC_primitive}, we get
\begin{equation}
 \lim\limits_{M \rightarrow \infty} \left(\tilde{ \pmb{R}} , D (1 + \Gamma D)^{M}  \Gamma^{\alpha,D} \pmb{P}_{\rm stat.}\right)=\lim\limits_{M \rightarrow \infty} \left(\tilde{ \pmb{R}}, D P_{\rm kMC}^M \Gamma^{\alpha,D} \pmb{P}_{\rm stat.}\right)=\left(\tilde{ \pmb{R}}, D \pmb{h}\pmb{b}'^T_0 \Gamma^{\alpha,D} \pmb{P}_{\rm stat.}\right).
\end{equation}
Now $\pmb{b}'^T_0 \Gamma^{\alpha,D} \pmb{P}_{\rm stat.}=(\pmb{b}'_0, \Gamma^{\alpha,D} \pmb{P}_{\rm stat.})=c$ is just a scalar and  eq. \ref{eq:h_vs_b} then leads to
\begin{equation}
 \lim\limits_{M \rightarrow \infty} \left(\tilde{ \pmb{R}} , D (1 + \Gamma D)^{M}  \Gamma^{\alpha,D} \pmb{P}_{\rm stat.}\right)=c\left(\tilde{ \pmb{R}}, D \pmb{h}\right) =c a \left(\tilde{ \pmb{R}},  \pmb{b}_0 \right) =0
\end{equation}
where we have used that $ \tilde{ {R}}_i=\frac{1}{\langle R \rangle} (R_i - \langle R \rangle)$ and thus $(\tilde{ \pmb{R}},  \pmb{b}_0)=0$.

\section{$\sum\limits_\alpha X^{\alpha}=1$}
We consider the stationary Master equation
\begin{equation}
 \Gamma \pmb{P}_{\rm stat.}=0.
\end{equation}
The generator $\Gamma$ depends linearly on the rate constants $k^\alpha$ (compare main manuscript). As we require $\pmb{P}_{\rm stat.}$ to be unique, the stationary distribution is a function of the rate constants $\pmb{P}_{\rm stat.}(\pmb k)$, where $\pmb k$ is the vector of all rate constants. If we now multiply $\pmb k$ with an arbitrary constant $\lambda\neq 0$, we find 
\begin{equation}
\begin{split}
 0&=\Gamma(\lambda \pmb k) \pmb{P}_{\rm stat.}(\lambda \pmb k)=\lambda\Gamma( \pmb k) \pmb{P}_{\rm stat.}(\lambda \pmb k) \\
 &\Rightarrow \Gamma( \pmb k) \pmb{P}_{\rm stat.}(\lambda \pmb k)=0.\\
 & \Rightarrow \pmb{P}_{\rm stat.}(\lambda \pmb k)=\pmb{P}_{\rm stat.}( \pmb k)
 \end{split}
\end{equation}
where we have used that the stationary distribution for a set of rate constants is unique. 
As the microscopic reaction rate $R_i$ depends linearly on the rate constants, we have for the average stationary reaction
\begin{equation}
 \langle R \rangle (\lambda \pmb k) = (\pmb R (\lambda \pmb k),\pmb{P}_{\rm stat.}(\lambda \pmb k))=\lambda(\pmb R ( \pmb k),\pmb{P}_{\rm stat.}( \pmb k))=\lambda\langle R \rangle (\pmb k),
\end{equation}
i.e.  $\langle R \rangle$ is homogeneous function of grade one of the rate constants. As long  $\langle R \rangle ( \pmb k)$ is differentiable, it holds
\begin{equation}
 \langle R \rangle ( \pmb k)=\sum\limits_\alpha \frac{\partial \langle R \rangle ( \pmb k)}{\partial k^\alpha} k^\alpha \quad \Rightarrow \quad 1=\sum\limits_\alpha \frac{\partial \langle R \rangle ( \pmb k)}{\partial k^\alpha} \frac{k^\alpha}{\langle R \rangle ( \pmb k)}=\sum\limits_\alpha X^{\alpha} 
\end{equation}
where we used the definition of the Degree of Rate Sensitivity $ X^{\alpha} := \frac{k^\alpha}{\langle R \rangle }\frac{\partial \langle R \rangle }{\partial k^\alpha} $. The sum rule also holds for the more common Degree of Rate Control\cite{campbell1994future}, because it is is simply the sum of the Degrees of Rate Sensitivity for forward and the backward reaction.

\section{Reaction rates for the CO oxidation on the RuO$_2$ surface}
\begin{figure}
\includegraphics[width=\linewidth]{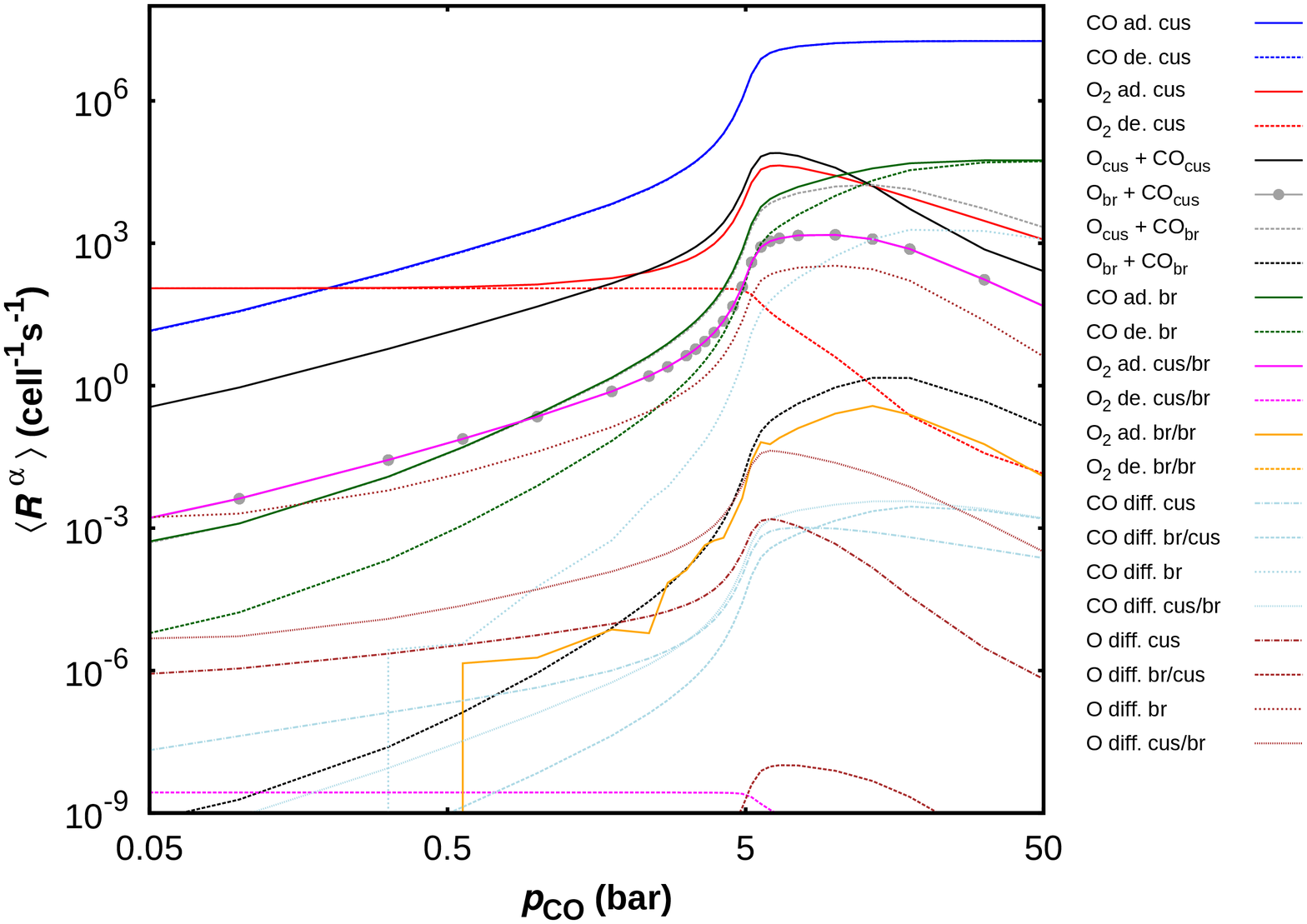}
\caption{The sampled TOFs for the elementary reactions as function of the CO partial pressure $p_{\text{CO}}$ ($T$=600\,K, $p_{\text{O}_2}=1$\,bar).
\label{all_tofs_pCO}}
\end{figure}
Figure \ref{all_tofs_pCO} display the stationary reaction rates for all elementary reactions  for the CO oxidation on the RuO$_2$(110) surface and the reaction conditions $T=600$\,K, $p_{\text{O}_2}=1$\,bar, and varying $p_{\text{CO}} \in [0.05,50]$\,bar. The results have been obtained from kinetic Monte Carlo simulation using the model by Reuter\cite{reuter2006first}. We employed a lattice of $20\times 20$ units cells and time-averages over $10^9$ steps after an initial relaxation of $10^8$ steps. Units are in number of reactions per surface unit cell and second.

\section{Periodic summation tree}

The periodic summation tree is data structure to sum up the last $M$ steps during the kinetic Monte Carlo sampling. For our purposes, we can restrict to perfect binary trees, i.e. $M$ is a power of $2$ 
and the depth of all leaves is equal. Leaves carry the values to be summed up from the last $M$ steps and all other nodes carry the sum of its children. The top node therefore carries the desired sum. If we advance one step, we only need to exchange the value stored at the oldest step and update only its ancestors. 
The tree itself can be stored in an one-dimensional array with $2M-1$ elements: the first entry stores the value at the top node, the next two the values at its children, the following four entries store the values at the following child nodes and so on until the last $M$ entries store the history. When the data is initially arranged in temporal order, we can create the periodic update by just storing the position of the oldest step and changing this once we have added a new leave value. The update algorithm is then 

\begin{algorithmic}
	\Procedure {updateTree}{$tree$,$oldest$, $newLeave$}
		\State $tree(oldest)\gets newLeave$
		\State $pos\gets oldest$
		\While{$pos>1$}
			\State $sibling \gets pos -2\cdot \mathrm{ mod }(pos,2) +1$
			\State $parent \gets \lfloor\frac{pos}2 \rfloor$
			\State $tree(parent) \gets tree(pos)+tree(sibling)$
			\State $pos\gets parent$
		\EndWhile 
		\State $oldest\gets \mathrm{mod}(((oldest+1)-M),M)+M$
	\EndProcedure
\end{algorithmic}
where $tree$ is the array storing the node values, $oldest$ is the position of the oldest step, and $newLeave$ is the new value, which shall be added to the tree.
Note, that we use the Fortran index convention, i.e. the first array element is indexed with $1$.

\section{TAC and IRF sampling}
The periodic summation tree is used for sampling the time-integrated auto-correlation (TAC) and the integrated response functions (IRF).
We will consider the estimation of both from a kinetic Monte Carlo trajectory of $N+1$ steps and using a truncation limit $M$, for which we assume $N>M$.

For the sampling of the TAC $c_R$ from a sequence $\{i_0,\ldots \i_N\}$ of states, we need to perform the summation
\begin{equation}\label{eq.TACEstimate2}
 c_R\approx  \frac{2}{T} \sum\limits_{n=0}^{N-M}\sum\limits_{l=0}^{M}  \delta R_{i_{l+n}}\Delta t_{l+n}  \delta R_{i_{n}} \Delta t_{n}
\end{equation}
where $T=\sum\limits_{l=0}^{N-M} \Delta t_l$ and $\Delta t_l$ is a (pre-averaged) time step. $\delta R_i=R_i -\langle R \rangle$ is the deviation of the microscopic reaction rate from its stationary average. With the definition
\begin{equation}
 z_{n}:=\begin{cases}
                 0 &\text{if} ~ n < 0 ~\text{or}~ n > N-M\\
                 1 & \text{else}
                \end{cases} .
\end{equation}
we can rewrite Eq. \ref{eq.TACEstimate2} as
\begin{equation}\label{eq.TACEstimate3}
 c_R\approx  \frac{2}{T} \sum\limits_{n=0}^{N}  \delta R_{i_{n}}\Delta t_{n} \sum\limits_{l=0}^{M} \delta R_{i_{n-l}} \Delta t_{n-l} z_{n-l}
\end{equation}
This we can calculate on the fly, i.e. during the generation of $\{i_0,\ldots \i_N\}$. The corresponding algorithm is
\begin{algorithmic}
	\Procedure {tacSampling}{$N$}
		\State $pos \gets M$ \Comment{position of  oldest step for the }
		\State $tree \gets 0$ \Comment{array for the tree}
		\State initialize the state $i$, time $T=0$, $c_R=0$
		\For{$n= 0:N$}
			\State $\Delta t \gets 1/{ w^{{\rm acc.}}_i}$
			\State $T \gets T + z_n \Delta t$
			\State determine process $i\rightarrow j$
			\State $c_R \gets c_R + \delta R_i\Delta t (z_n \delta R_i\Delta t + tree(1))$
			\State \Call{updateTree}{$tree$,$pos$,$z_n \delta R_i\Delta t$ }
			\State $i \gets j$
		\EndFor
		\State $c_R \gets c_R/T$
	\EndProcedure.
\end{algorithmic}
For clarity we have only given the algorithm for a single value of $M$.

For the sampling of the IRF $X^\alpha_1$ with respect to the reaction $\alpha$, the corresponding equation to Eq. \ref{eq.TACEstimate2}  is
\begin{equation}\label{eq:IRFEstimator2}
\begin{split}
 X^\alpha_1 & \approx \frac{1}{T} \sum\limits_{n=0}^{N-M-1}\Bigg[ - \tilde R_{i_{n}} w^{\alpha,{\rm acc.}}_{i_{n}}\Delta t_n^2 \\
 &+ \sum\limits_{l=1}^{M+1} \tilde R_{i_{n+l}}\Delta t_{n+l} (O^\alpha_{i_{n+1}i_{n}}- w^{\alpha,{\rm acc.}}_{i_{n}}) \Delta t_{n} \Bigg],
 \end{split}
\end{equation}
where $T=\sum\limits_{l=0}^{N-M-1} \Delta t_l$ and $\tilde R_{i}=\delta R_i/\langle R \rangle$. $w^{\alpha,{\rm acc.}}_{i}$ is the accumulated rate of the reaction $\alpha$ and
\begin{equation}
 O^\alpha_{ij}= \begin{cases}
                 w^{{\rm acc.}}_j &\text{if} ~ j\rightarrow i \in \alpha\\
                 0 & \text{else}
                \end{cases} .
\end{equation}
with the total accumulated rate $w^{{\rm acc.}}_j$. We now redefine $z_n$ 
\begin{equation}
 z_{n}:=\begin{cases}
                 0 &\text{if} ~ n < 0 ~\text{or}~ n > N-M-1\\
                 1 & \text{else}
                \end{cases} .
\end{equation}
This allow us to rewrite Eq. \ref{eq:IRFEstimator2}
\begin{equation}\label{eq:IRFEstimator3}
\begin{split}
X^\alpha_1 & \approx \frac{1}{T} \sum\limits_{n=0}^{N}\Bigg[ - \tilde R_{i_{n}} w^{\alpha,{\rm acc.}}_{i_{n}}\Delta t_n^2 z_n \\
 &+ \sum\limits_{l=1}^{M+1} \tilde R_{i_{n}}\Delta t_{n} (O^\alpha_{i_{n-l+1}i_{n-l}}- w^{\alpha,{\rm acc.}}_{i_{n-l}}) \Delta t_{n-l}z_{n-l} \Bigg].
 \end{split}
\end{equation}
The algorithm for performing the summation on the fly is very similar to that for estimating the TAC. We only provide the pseudocode for a single $\alpha$ and only a single value of $M$.
\begin{algorithmic}
	\Procedure {irfSampling}{$N$}
		\State $pos \gets M$ \Comment{position of  oldest step for the }
		\State $tree \gets 0$ \Comment{array for the tree}
		\State initialize the state $i$, time $T=0$, $X^\alpha_1=0$
		\For{$n= 0:N$}
			\State $\Delta t \gets 1/{ w^{{\rm acc.}}_i}$
			\State $T \gets T + z_n \Delta t$
			\State determine process $i\rightarrow j$

			\State $X^\alpha_1=X^\alpha_1 + \tilde R_{i}\Delta t( -w^{\alpha,{\rm acc.}}_{i}\Delta t z_n + tree(1))$
			\State $O \gets  -w^{\alpha,{\rm acc.}}_{i}\Delta t z_n$
			\If {$i\rightarrow j \in \alpha$} 
				\State $O\gets O + w^{{\rm acc.}}_i\Delta tz_n$
			\EndIf
			\State \Call{updateTree}{$tree$,$pos$,$O$}
			\State $i \gets j$
		\EndFor
		\State $X^\alpha_1=X^\alpha_1/T$
	\EndProcedure
\end{algorithmic}

%
\end{document}